\documentclass[aps,pra,reprint,
showkeys,
letterpaper,
superscriptaddress]{revtex4-2}%

\usepackage{amsfonts}
\usepackage{amsmath}
\usepackage{amssymb}
\usepackage{graphicx}%

\usepackage[pdftex,
pdfa,
hyperindex=true, unicode, bookmarks=true, pdfpagelabels=true]{hyperref}
\usepackage[figure,figure*,table,table*]{hypcap}
\usepackage[hypcap=true]{caption}
\usepackage{orcidlink}
 \definecolor{green}{rgb}{0,0.5,0}
\hypersetup{colorlinks=true,urlcolor=blue,linkcolor=green}
\hypersetup{
pdfauthor={Szil\'{a}rd Sajti, L\'{a}szl\'{o} De\'{a}k},
pdftitle={Anisotropic extension of the Parratt formalism}}

\def\d{\,\mathrm{d}} 
\def\e{\,\mathrm{e}} 
\def\imunit{\mathrm{i}} 

\newcommand{\tens}[1]{\overline{\overline{\mathsf{#1}}}}
\newcommand{\tenscr}[1]{\overline{\overline{\mathcal{#1}}}}
\newcommand{\tensfr}[1]{\overline{\overline{\mathfrak{#1}}}}
\newcommand{\tenshat}[1]{\hat{\mathsf{#1}}}
\newcommand{\tensvecg}[1]{\overline{\overline{\boldsymbol{#1}}}}

\newcommand{\vect}[1]{\mathbf{#1}}

\def\irott#1{\mathcal{#1}}
\def\irottv#1{\underline{\irott{#1}}}

\newcommand{\Imag}{\mathop{\mathrm{Im}}} 

\def\Laplace{\text{\rotatebox[origin=c,units=360]{180}{\(\nabla\)}}}

\newcommand{\Nquad}{\hspace{0.25em}} 

\newtheorem{theorem}{Theorem}

\newtheorem{dataavailability}[theorem]{Data availability}

\begin{document}
\title{Anisotropic extension of the Parratt formalism}
\email{sajti.szilard@wigner.hun-ren.hu}
\author{\orcidlinki{Szil\'{a}rd Sajti}{0000-0002-8748-8242}}
\affiliation{Department of Materials Science by Nuclear Methods, Institute for Particle and Nuclear Physics, HUN-REN Wigner Research Centre for Physics, P.O.\ Box 49, Budapest, H-1525, Hungary}
\author{\orcidlinki{L\'{a}szl\'{o} De\'{a}k}{0000-0002-1558-2109}}
\affiliation{Department of Materials Science by Nuclear Methods, Institute for Particle and Nuclear Physics, HUN-REN Wigner Research Centre for Physics, P.O.\ Box 49, Budapest, H-1525, Hungary}
\affiliation{Department of Natural Science, Ludovika University of Public Service, P.O.\ Box 60, Budapest, H-1441, Hungary}
\pacs{PACS number}

\begin{abstract}
Neutron and X-ray reflectometry are important methods for studying thin multilayer systems. The Parratt method and the method of characteristic matrices, also referred to as transfer matrices, are used for simulation, evaluation of experimental results, and designing optical systems, like mirrors. The Parratt method had been derived for isotropic systems. The method of characteristic matrices can also handle anisotropic problems, but it is burdened with numerical instabilities, which may arise in the case of thick samples at grazing angle incidence.

In this paper, we derive a generalized Parratt method applicable to anisotropic systems. Furthermore, as we show, this is devoid of the numerical instabilities arising in the method of characteristic matrices. We derive formulae for both reflectivity and transmissivity. The stability of the new approach is demonstrated by comparing calculated results obtained via different methods. The problem of rough interfaces is also addressed, and the results gained by different approximations for some systems are compared.
\end{abstract}

%
%
%

\maketitle

\section{Introduction}\label{sec:Intro}
Reflectometric methods reveal the one-dimensional scattering amplitude density profile normal to a surface. This profile is directly linked to variations in chemical composition, isotopic ratios, magnetic properties, and other characteristics within the depth the radiation penetrates. Consequently, X-ray and neutron reflectometry are widely used to characterize surfaces, thin films, and layered materials \cite{Merkel-2011, Merkel-2019, Lengyel-2022, Merkel-2022, Perissinotto-2019, Perissinotto-2021}. For non-resonant X-rays or unpolarized neutrons (and non-magnetic samples), the scattering behavior is independent of wave polarization, allowing the layered structure to be modeled using a single complex refractive index. However, when the material exhibits birefringence, polarization-dependent scattering occurs, requiring a more complex, non-scalar optical treatment. In Ref.~\cite{Deak01,DeakPhd}, it has been shown that a common optical formalism exists for the
anisotropic neutron and anisotropic nuclear resonant X-ray transmission and
reflection for the case of forward scattering and that of grazing incidence.

We frequently use transfer matrix methods \cite{Abeles-1950, BornWolf, Deak96, Ruhm99, RohlsbergerHypInt99, Deak01, Daillant-b2009}, also known as the method of
characteristic matrices, to evaluate the results of such experiments or to design devices based on these systems. These methods exhibit numerical
instability for large systems containing many layers because of exponentially growing terms \cite{Ko-Sambles-1988, Moharam-1995,
LifengLi-1993, LifengLi-1994, Ning-Tan-2009, Macke-2014}. Parratt in Ref.~\cite{Parratt-1954} uses a different recursive approach. This
is numerically stable according to experience, but it was derived for isotropic layers. In this paper, we demonstrate how this method can be modified
for anisotropic layers, in such a way that it is devoid of the numerical instabilities arising from the exponential terms of the transfer matrices.
These cases include polarized neutron reflectometry (PNR) and (synchrotron) M\"{o}ssbauer reflectometry (SMR), the latter is only a special but well-studied case of
the anisotropic (resonant) x-ray scattering problem.

In the next section, we outline the method of characteristic matrices in the form of Ref.\ \cite{Deak01} and the basic theoretical
assumptions. We briefly review the Parratt method, published in Ref.\ \cite{Parratt-1954}, in section \ref{sec:OrigParrattForm}.  We derive a generalized Parratt method from the method of characteristic matrices applicable to anisotropic systems in section \ref{sec:ParrMethCharMatr}. In section \ref{sec:ParrattTransfMatr}, we will introduce another transfer matrix method and another generalized Parratt method derived from that. This latter generalized Parratt method is the one we implemented and compared with the results of the first transfer matrix method, i.e., the method of characteristic matrices, in section \ref{sec:ExamplesCompar}. Furthermore, we derive a transmission formula and two analytical approximations for the generalized Parratt method, taking into account the roughness of the interfaces as well, in section \ref{sec:ParrattTransfMatr}.

\section{Method of characteristic matrices}\label{sec:MethCharMatr}

In this section, starting from the general theory of Lax \cite{Lax51}, we
shall repeat some general formulae for the scattering of multicomponent waves.
Our earlier results are reformulated to make them suitable for applying the method to a generalisation of the Parratt method.

For many scattering centers, the coherent field $\Psi\left(  \mathbf{r}\right)
$, representing an electromagnetic field vector or quantum mechanical spinor
state, fulfills the%

\begin{equation}
\left[  \left(  \Delta+k^{2}\right)  \tens{I}+4\pi N\,\tens{f}\right]  \Psi\left(
\mathbf{r}\right)  =\mathbf{0,} \label{eq:Lax_3}%
\end{equation}
three-dimensional wave equation, where $k$ is the vacuum wavenumber, $\tens{I}$ is
the $2\times2$ unit matrix, $\tens{f}$ is the coherent forward scattering amplitude,
$N$ is the density of the scattering centers per unit volume, and $\Psi\left(
\mathbf{r}\right)  $ is the coherent field defined by an average of the field
vectors over the positions and states of the scattering centers \cite{Lax51}.
Eq.~(\ref{eq:Lax_3}) shows that from the point of view of the coherent field, the
system of randomly distributed scattering centers can be replaced by a
homogeneous medium, with an index of refraction $n=I+\frac{2\pi N}{k^{2}}f,$ where it is exploited that
$n$ hardly differs from $I$ for both x-rays and slow neutrons. It is often
practical to use the tensor defined by $\tens{\chi}=$ $\frac{4\pi
N}{k^{2}}\tens{f}$ \cite{Deak96}, which is the susceptibility in the case of X-rays (represented by a \(2\times2\) complex matrix), and
\begin{equation}
\tens{\chi}=\frac{1}{k^2}\left[\frac{2m_{n^0}}{\hbar^2} g\mu_\mathrm{N} \vect{B}_\text{eff}\tensvecg{\sigma}-4\pi s\tens{I}\right]
\end{equation}
 for neutrons, where \(s\) is the scattering length density, \(\vect{B}_\text{eff}\) is the effective magnetic induction, \(m_{n^0}\) is the neutron mass, \( g\) is the Landé g-factor, \(\mu_\mathrm{N}\) is the nuclear magneton. In the following, we will also use the name susceptibility tensor in the case of neutrons for practical reasons.

Thus, the wave equation for X-ray and neutron reflectometry can be written formally in the same form \cite{Deak01}:
\begin{equation}
 \left[\Laplace + k^2\left(\tens{I} +\tens{\chi}\right)\right]\Psi(\vect{r}) = 0. \label{eq:WaveEquationChiCommon}
\end{equation}

By choosing a simple homogeneous layer with the above susceptibility $\chi$
and $z$ axis normal to the layer, one gets the 1D wave equation:
\begin{equation}
\d^2_z\Psi\left(  z\right)  +k^{2}\sin\theta\left[  \tens{I}\sin\theta
+\frac{\tens{\chi}}{\sin\theta}\right]  \Psi\left(  z\right)  =\mathbf{0,}
\label{eq:base1}%
\end{equation}
with $\theta$ being the grazing angle of incidence. Defining $\Phi$ via $\left(
\imunit k\sin\theta\right)  \, \d_z \Phi\left(  z\right)  := \d^2_z\Psi\left(  z\right)  $, we get a system of first-order differential equations:
\begin{equation}
\frac{\mathrm{d}}{\mathrm{d}z}\left(
\begin{array}
[c]{c}%
\Phi\\
\Psi
\end{array}
\right)  = \imunit k\tens{M}\left(
\begin{array}
[c]{c}%
\Phi\\
\Psi
\end{array}
\right)  , \label{eq:Neuform}%
\end{equation}
where
\begin{equation}
\tens{M}=\left(
\begin{array}
[c]{cc}%
0 & \tens{I}\sin\theta+\frac{\tens{\chi}}{\sin\theta}\\
\tens{I}\sin\theta & 0
\end{array}
\right)  \label{eq:basicM}%
\end{equation}
is usually called the \textquotedblright differential propagation
matrix\textquotedblright\ in optics \cite{Deak96}. Eq.~(\ref{eq:Neuform}) was
derived without specifying the scattering process.

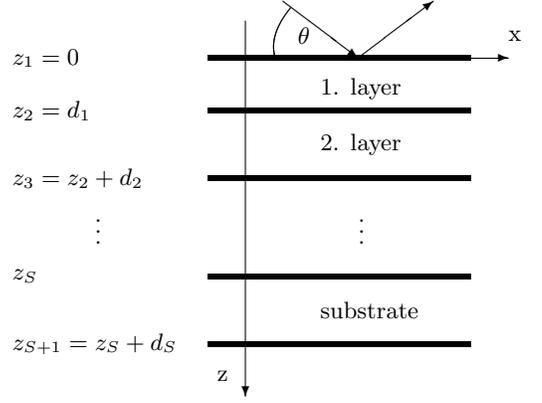
\begin{figure}[!ht]
\unitlength 1mm
\begin{picture}(70,55)(0,0)
\thinlines
\put(28,45){\vector(1,0){40}} \put(68,48){\makebox(0,0)[l]{x}}
\put(33,50){\vector(0,-1){50}}\put(30,2){\makebox(0,0)[b]{z}}

\qbezier(37,45)(36,48)(39,51.7)
\linethickness{1mm}
\put(40,48){\makebox(0,0)[l]{$\theta$}}
\put(38,52.6){\vector(4,-3){10}}
\put(48.,45.1){\vector(4,3){10}}

\linethickness{.75mm}
\put(28,45){\line(1,0){35}} \put(2,45){\makebox(0,0)[l]{\(z_1 = 0\)}}
\put(43,41){\makebox(0,0)[l]{1. layer}}
\put(28,38){\line(1,0){35}} \put(2,38){\makebox(0,0)[l]{\(z_2 = d_1\)}}
\put(43,33.5){\makebox(0,0)[l]{2. layer}}
\put(28,29){\line(1,0){35}} \put(2,29){\makebox(0,0)[l]{\(z_3 = z_2+d_2\)}}
\put(13,23){\makebox(0,0)[l]{\(\vdots\)}}   \put(48,23){\makebox(0,0)[l]{\(\vdots\)}}
\put(28,16){\line(1,0){35}} \put(2,16){\makebox(0,0)[l]{\(z_S \)}}
\put(43,11.5){\makebox(0,0)[l]{substrate}}
\put(28,7){\line(1,0){35}} \put(2,7){\makebox(0,0)[l]{\(z_{S+1} = z_S + d_S\)}}
\end{picture}
\caption{Multilayer sample and coordinate-system}
\end{figure}
For an arbitrary multilayered film with homogeneous layers of thicknesses
$d_{1},d_{2}, \dots, d_{S-1}$ and differential propagation matrices $\tens{M}_{1}%
,\tens{M}_{2}, \dots, \tens{M}_{S-1},$ $\tens{\chi}$ in Eq. (\ref{eq:basicM}) is replaced by the
susceptibility $\tens{\chi}_{l}$ of layer $l$ $(l=1, \dots, S-1)$. We have reserved $l=0$
for the semi-infinite medium of the incident radiation, and $l=S$ for the
substrate layer, which we also regard as a semi-infinite medium in the case of reflectometry. The solution of the differential equation (\ref{eq:Neuform})
can be expressed in terms of the total characteristic matrix
\begin{equation}
\tens{L}=\tens{L}_{S-1}\cdot \ldots \cdot \tens{L}_{2}\cdot \tens{L}_{1} \label{eq:Ltotal}%
\end{equation}
of the multilayer, where
\begin{equation}
\tens{L}_{l}=\exp\left(  \imunit kd_{l}\tens{M}_{l}\right)  \label{eq:basicL}%
\end{equation}
is the characteristic matrix of the $l^{\text{th}}$ individual layer. Note
that the solution of the linear first-order differential Eq. (\ref{eq:Neuform})
means that in a medium $l$ described by the characteristic matrix $\tens{L}_{l},$ the
fields at depth $z=0$ are connected by the fields at depth $z$ by the
relation
\begin{equation}
\left(
\begin{array}
[c]{c}%
\Phi\\
\Psi
\end{array}
\right)  \left(  z\right)  =\tens{L}_{l}\left(
\begin{array}
[c]{c}%
\Phi\\
\Psi
\end{array}
\right)  \left(  0\right)  , \label{eq:NNeuform}%
\end{equation}
and the coherent field $\Psi$, as well as the derived field $\Phi$ must be
continuous at each boundary.

The $2\times2$ reflectivity matrix $\tens{R}$ is calculated from the total
characteristic matrix $\tens{L}$ by
\begin{eqnarray}
\tens{R}&=\left[  \left(  \tens{L}_{\left[  11\right]  }-\tens{\gamma}_{S}\tens{L}_{\left[  21\right]
}\right)  \tens{\gamma}_{0}-\tens{L}_{\left[  12\right]  }+\tens{\gamma}_{S}\tens{L}_{\left[  22\right]
}\right]  ^{-1}\nonumber\\
&\left[  \left(  \tens{L}_{\left[  11\right]  }-\tens{\gamma}_{S}\tens{L}_{\left[
21\right]  }\right)  \tens{\gamma}_{0}+\tens{L}_{\left[  12\right]  }-\tens{\gamma}_{S}\tens{L}_{\left[
22\right]  }\right]  , \label{eq:rDeak}%
\end{eqnarray}
where $\tens{L}_{\left[  ij\right]  }$ $\left(  i,j=1,2\right)  $ are $2\times2$
blocks of the $4\times4$ total characteristic matrix $\tens{L}$ \cite{Deak96}, and
the impedance tensors $\tens{\gamma}_{l}^{0,r}$ were introduced by the relationship:
\begin{equation}
\tens{\gamma}_{l}^{0,r}\Psi_{l}^{0,r}:=\Phi_{l}^{0,r}, \label{eq:gammaNeu}%
\end{equation}
where indexes $0$ and $r$ indicate incident and reflected waves, in layer
(medium) $l$, respectively (see Eq. (3.4) of Ref.~\cite{Deak96}). Substituting
the definition (\ref{eq:gammaNeu}) back into Eq. (\ref{eq:Neuform}), using Eq.
(\ref{eq:basicM}), the impedance tensor $\tens{\gamma}_{l}^{0,r}$ reads%

\begin{equation}
\tens{\gamma}_{l}^{0}=-\tens{\gamma}_{l}^{r}=\sqrt{\tens{I}+\frac{\tens{\chi}_{l}}{\sin^{2}\theta}},
\label{eq:gamma0r}%
\end{equation}
for an incident (index $0$) and reflected (index $r$) wave in layer $l$,
respectively. Based on the simple relation (\ref{eq:gamma0r}), the upper index is
dropped, and we use $\tens{\gamma}_{l}$ instead of $\tens{\gamma}_{l}^{0}$ and
$-\tens{\gamma}_{l}^{r}$ for any medium $l=0, \dots, S$.

The reflected intensity reads%
\begin{equation}
I^{r}=\mathrm{Tr}\left(  \tens{R}^{H}\tens{\rho}_{\text{a}}\tens{R}\tens{\rho}_{\text{p}}\right)  ,
\label{eq:ReflInt}%
\end{equation}
which can be calculated using an arbitrary polarization density matrix
$\tens{\rho}_{\text{p}}$ for the incident beam (polariser), $\tens{\rho}_{\text{a}}$ for the
outgoing beam (analyzer), and the reflectivity matrix $\tens{R}$ \cite{Blume68}. If we
have only one interface between the initial ($l=0)$ and the substrate ($l=S$)
media, the multilayer can be interpreted as $d_{l}=0$ \ (for each
$l=1, \dots, S-1)$, therefore Eq. (\ref{eq:basicL}) gives $\tens{L}_{\left[  11\right]
}=\tens{L}_{\left[  22\right]  }=\tens{I}$ and $\tens{L}_{\left[  12\right]  }=\tens{L}_{\left[
21\right]  }=0$, finally the reflectivity matrix (\ref{eq:rDeak}) reads%
\begin{equation}
\tens{r}=\left[  \tens{\gamma}_{0}+\tens{\gamma}_{S}\right]  ^{-1}\left[  \tens{\gamma}_{0}-\tens{\gamma}
_{S}\right]  . \label{aniFresnel0S}%
\end{equation}
which are the Fresnel coefficients in the anisotropic case, calculated for the
interface situated between $l=0$ (initial medium) and $l=S$ (substrate). This
Fresnel reflectivity matrix for the interface between layers $l-1$ and $l$
reads%
\begin{equation}
\tens{r}_{l}=\left[  \tens{\gamma}_{l-1}+\tens{\gamma}_{l}\right]  ^{-1}\left[  \tens{\gamma}_{l-1}%
-\tens{\gamma}_{l}\right]  . \label{aniFresnel_l}%
\end{equation}

The exponential of the $4\times4$ matrix in Eq. (\ref{eq:basicL}) was given in
\cite{Deak01}, as
\begin{equation}
\tens{L}_{l}=\left(
\begin{array}
[c]{cc}%
\cosh\left(  \imunit kd_{l}\tens{\gamma}_{l}\sin\theta\right)  & \tens{\gamma}_{l}\sinh\left(
\imunit kd_{l}\tens{\gamma}_{l}\sin\theta\right) \\
\tens{\gamma}_{l}^{-1}\sinh\left(  \imunit kd_{l}\tens{\gamma}_{l}\sin\theta\right)  & \cosh\left(
\imunit kd_{l}\tens{\gamma}_{l}\sin\theta\right)
\end{array}
\right)  , \label{eq:LDeak}%
\end{equation}
where the impedance tensor $\tens{\gamma}_{l}$ appears in the matrix elements,
and we note that the arguments of the $\sinh$ and $\cosh$ functions are
$2\times2$ matrices.

\section{Parratt formalism}\label{sec:OrigParrattForm}

According to the original work of Parratt \cite{Parratt-1954}, the reflection coefficient on the interface between the \(n-1\)-th and \(n\)-th layer is
\begin{equation}
R_{n-1,n} = a_{n-1}^4 \frac{F_{n-1,n} + R_{n,n+1}}{1 + F_{n-1,n}R_{n,n+1}}, \label{eq:Parratt}
\end{equation}
where \(F_{n-1,n} = \frac{g_{n-1}-g_{n}}{g_{n-1}+g_{n}}\) is the Fresnel reflection coefficient between these interfaces. The amplitude factor \(a_n = \e^{-\imunit k g_n \frac{d_n}{2}}\) depends on the layer thickness \(d_n\) and on the sample plane perpendicular component of the wavenumber \(k g_n.\)

In X-ray (and neutron) reflectometry, we usually assume that the beam is absorbed in the substrate. Therefore, it does not reach the lower interface of the substrate. Thus, the last reflecting interface is the top interface of the substrate. Nevertheless, this is not always true. E.g.: transmission experiments occur at grazing incidence for neutrons \cite{Ott-2011} as well. If the last reflecting interface is between the layers with indices \((\Omega,\Omega+1),\) then there is no reflection from the next interface  \(R_{\Omega+1,\Omega+2} = 0,\) thus according to (\ref{eq:Parratt}) \(R_{\Omega,\Omega+1} = a_{\Omega}^4 F_{\Omega,\Omega+1}.\) This is used as the initial value of the recurrence relation.

\section{Derivation of the anisotropic Parratt formula}\label{sec:ParrMethCharMatr}

After the previous section, we are ready to derive the Parratt equation
\cite{Parratt-1954} for the anisotropic case from the transfer matrix method reviewed in section \ref{sec:MethCharMatr}. In the first Parratt iteration step, the
reflectivity is calculated for the interface of two semi-infinite media; in
fact, the Fresnel reflection coefficient is calculated according to
(\ref{aniFresnel_l}) for the interface $S$, between layers $l=S-1$ and
$\ l=S,$ i.e., the top interface of layer $l$ is
also noted as interface $l$. The reflectivity in the first iteration step is%
\begin{equation}
\tens{R}_{S}=\tens{r}_{S}=\left[  \tens{\gamma}_{S-1}+\tens{\gamma}_{S}\right]  ^{-1}\left[  \tens{\gamma}
_{S-1}-\tens{\gamma}_{S}\right]  . \label{Parratt_it1}%
\end{equation}
The first iteration step is special, as there is no reflected wave in the
substrate layer; only a refracted wave occurs. From the second iteration step
onwards, both the incoming and outgoing waves appear on both the lower and the upper interfaces, therefore both fields $\Psi_{l}$, $\Phi_{l}$ have
non-zero incoming $\Psi_{l}^{0}$, $\Phi_{l}^{0}$ and outgoing $\Psi_{l}^{r}$
and $\Phi_{l}^{r}$  components in the neighboring layers, which are superposed and read for any $l=2,\dots,S$, as
\[
\left(
\begin{array}
[c]{c}%
\Phi_{l}\\
\Psi_{l}%
\end{array}
\right)  =\left(
\begin{array}
[c]{c}%
\Phi_{l}^{0}+\Phi_{l}^{r}\\
\Psi_{l}^{0}+\Psi_{l}^{r}%
\end{array}
\right)  =\left(
\begin{array}
[c]{c}%
\tens{\gamma}_{l}\left(  \tens{I}-\tens{R}_{l}\right)  \Psi_{l}^{0}\\
\left(  \tens{I}+\tens{R}_{l}\right)  \Psi_{l}^{0}%
\end{array}
\right)
\]
at the bottom interface of layer $l$, and similarly, at the top interface, we have to replace all the indices \(l\) with \((l-1).\)
The top and bottom interfaces are now connected by the characteristic matrix
$\tens{L}_{l}$ of layer $l$ according to (\ref{eq:NNeuform})%
\begin{equation}
\left(
\begin{array}
[c]{c}%
\tens{\gamma}_{l}\left(  \tens{I}-\tens{R}_{l}\right)  \Psi_{l}^{0}\\
\left(  \tens{I}+\tens{R}_{l}\right)  \Psi_{l}^{0}%
\end{array}
\right)  =\tens{L}_{l}\left(
\begin{array}
[c]{c}%
\tens{\gamma}_{l-1}\left(  \tens{I}-\tens{R}_{l-1}\right)  \Psi_{l-1}^{0}\\
\left(  \tens{I}+\tens{R}_{l-1}\right)  \Psi_{l-1}^{0}%
\end{array}
\right)  , \label{parratt_base}%
\end{equation}
where the $4\times4$ matrix equation contains the relation of the unknown
reflectivity $\tens{R}_{l-1}$ at the upper interface to the already known $\tens{R}_{l}$
reflectivity of the lower interface. Indeed, Eq. (\ref{parratt_base}) is the
iteration step identical with the Parratt formalism, but given for the
anisotropic case. To prove that statement, one can express $\Psi_{l}^{0}$ from
the second line of the matrix equation (\ref{parratt_base}), and replace that
to its first line, using also Eq. (\ref{eq:LDeak}), so the upper reflectivity
$\tens{R}_{l-1}$ reads (for further details see Appendix \ref{app:DeakParratt})%
\begin{equation}
\tens{R}_{l-1}=\left(  \tens{I} {- \tens{r}_{l}}\right)  \left[  {\widehat{R}
_{l}\tens{r}_{l}} +\tens{I}\right]  ^{-1}\left[  \widehat{R}_{l}+\tens{r}_{l}\right]  \left(
\tens{I} {-\tens{r}_{l}}\right)  ^{-1}, \label{deakParratt}%
\end{equation}
where $\tens{r}_{l}$ is the Fresnel coefficient at the interface $l$, as given in Eq.
(\ref{aniFresnel_l}),
\begin{equation}
\widehat{R}_{l}=\exp\left(  \imunit kd_{l}\tens{\gamma}_{l}\sin\theta\right)  \tens{R}_{l}%
\exp\left(  \imunit kd_{l}\tens{\gamma}_{l}\sin\theta\right) \label{eq:RKalapDef}
\end{equation}
is the transformed reflectivity of $\tens{R}_{l}$. The matrix functions $\sinh\left(
\imunit kd_{l}\tens{\gamma}_{l}\sin\theta\right)  $ and $\cosh\left(  \imunit kd_{l}\tens{\gamma}
_{l}\sin\theta\right)  $ have been expressed in Eq.~(\ref{eq:LDeak}) by their
definitions given by the exponential matrices, as
 \(\sinh x =\frac{1}{2}[\exp(x)  -\exp(-x)],\) and \(\cosh x =\frac{1}{2}[\exp(x)  +\exp(-x)].\)
One has to be careful with the order of the calculation of the products in Eq.~(\ref{deakParratt}), because the operators $\tens{r}_l$, $\tens{\gamma}_{l},$ and $\tens{R}_{l}$ do not
commute in general.

In the case of isotropic layers, all the $2\times2$ matrices of Eq.~(\ref{deakParratt}) are scalar, hence all of the matrices $\tens{r}_{l},$
$\widehat{R}_{l},$ and $\tens{\gamma}_{l}$ commute with each other, and the
reflectivity matrix reads
\[
R_{l-1}=\frac{R_{l}\exp\left(  2ikd_{l}\gamma_{l}\sin\theta\right)  +r_{l}%
}{r_{l}R_{l}\exp\left(  2ikd_{l}\gamma_{l}\sin\theta\right)  +1},
\]
which expression is identical with the Parratt equation \cite{Parratt-1954}
being valid only for the isotropic case. One can easily show that \(k\gamma_l \sin \theta\) is identical with $k_{l}^{\perp},$ i.e., the
perpendicular component of the momentum vector in layer $l.$

During the second iteration step, the thickness of the layer labeled by $S-1$ is
also taken into account.

\section{A different transfer matrix approach, Parratt formalism for anisotropic layers}\label{sec:ParrattTransfMatr}

In this section, we employ an alternative approach to solve the wave equation (\ref{eq:WaveEquationChiCommon}) and obtain the Parratt formalism for anisotropic multilayers. This approach is chosen because it is well-suited to the derivation of the generalized Parratt method, enabling a clear separation of forward- and backward-propagating waves and providing insight into its stability. This method provides a slightly different and more detailed perspective; furthermore, it is the version used in our implementation.

As the plane parallel component (in the following denoted by \(\beta\)) of the wave vector is the same in each layer, we can assume the solution in the \(l\)-th homogeneous layer to be of the form
\begin{equation}
 \Psi(\vect{r}) = \psi_l(z) \e^{\imunit \beta x},\quad\text{where}
\end{equation}
\begin{eqnarray}
\psi_l(z) &= \irott{A}^{+}_{l;1} \vect{u}_{l;1} \e^{\imunit \kappa_{l;1} \zeta_l} + \irott{A}^{+}_{l;2} \vect{u}_{l;2} \e^{\imunit \kappa_{l;2} \zeta_l} + \nonumber\\
          &\irott{A}^{-}_{l;1} \vect{u}_{l;1} \e^{-\imunit \kappa_{l;1} \zeta_l} + \irott{A}^{-}_{l;2} \vect{u}_{l;2} \e^{-\imunit \kappa_{l;2} \zeta_l}, \label{eq:psilAlak}
\end{eqnarray}
where \(\zeta_l = z-z_l,\) and \(\kappa_{l;1}\) and \(\kappa_{l;2}\) are the roots of eigenvalues equation
\begin{equation}
\det \left[\left(-\kappa^2_l + k^2- \beta^2 \right) \tens{I} + k^2 \tens{\chi}_l \right] = 0,
\end{equation}
 and \(\vect{u}_{l;1}\) és \(\vect{u}_{l;2}\) are the corresponding 2-dimensional complex unit eigenvectors, which will be the eigenvectors of \(\tens{\chi}\) too. As in our systems, there is no amplification, therefore the sensible physical solutions should satisfy the condition \(\Imag \kappa_{l;\genfrac{}{}{0pt}{2}{1}{2}} \geqslant 0.\)
\(\irott{A}^{+}\) represents the amplitudes of the waves propagating in the direction of the substrate, and \(\irott{A}^{-}\) corresponds to the waves propagating in the opposite direction, i.e., the waves reflected in that direction. The numerical instabilities, as we will see later, arise from the members representing the backpropagating waves. These would not cause any problem if we could calculate with arbitrary precision. In numerical calculations, this is not the case; the finite number representation and the problems arising from this in the computation of the exponential of a matrix \cite{Moler-1978,Moler-2003} may lead to numerical instabilities.

The wave function \(\Psi\) and its first derivatives are continuous at the interfaces, thus
\begin{equation}
\left(\begin{array}{c}\psi_{l-1}(z_l) \\ \partial_z \psi_{l-1}(z_l)\end{array}\right) = \left(\begin{array}{c}\psi_{l}(z_l) \\ \partial_z \psi_{l}(z_l)\end{array} \right),
\end{equation}
which, using (\ref{eq:psilAlak}) and introducing the notations
\(\vect{A}_l = \left(\begin{array}{cccc} \irott{A}^+_{l;1}& \irott{A}^+_{l;2}& \irott{A}^-_{l;1}& \irott{A}^-_{l;2}\end{array}\right)^T\) and \(\irottv{A}^\pm_l = \left(\begin{array}{cc} \irott{A}^\pm_{l;1}& \irott{A}^\pm_{l;2}\end{array}\right)^T,\) can be written in the form
\begin{equation}
\tens{W}_{l-1}(d_{l-1}) \vect{A}_{l-1}
= \tens{W}_l(0) \vect{A}_l,\label{eq:WMatrixDef}
\end{equation}
where \(d_l\) is the thickness of the \(l\)-th layer, and \(\tens{W}_{l}\) is
\begin{widetext}
\begin{eqnarray}
\tens{W}_{l}\left(\zeta_l\right) &=&
 \tens{W}_{l}(0) \tenscr{P}_{l}\left(\zeta_l\right)=
\left(\begin{array}{cccc}  \vect{u}_{l;1} \e^{\imunit \kappa_{l;1} \zeta_l} & \vect{u}_{l;2} \e^{\imunit \kappa_{l;2} \zeta_l} & \vect{u}_{l;1} \e^{-\imunit \kappa_{l;1} \zeta_l} & \vect{u}_{l;2} \e^{-\imunit \kappa_{l;2} \zeta_l}\\
\imunit \kappa_{l;1} \vect{u}_{l;1} \e^{\imunit \kappa_{l;1} \zeta_l} & \imunit \kappa_{l;2} \vect{u}_{l;2} \e^{\imunit \kappa_{l;2} \zeta_l} & -\imunit \kappa_{l;1} \vect{u}_{l;1} \e^{-\imunit \kappa_{l;1} \zeta_l} & -\imunit \kappa_{l;2} \vect{u}_{l;2} \e^{-\imunit \kappa_{l;2} \zeta_l}
\end{array}\right),\quad\text{
where}\quad\label{eq:PHullTranszferMatrixDiag}\\
\tenscr{P}_{l}\left(\zeta_l\right) &=& \left(\begin{array}{cc}
\tenscr{P}_{l}^+\left(\zeta_l\right)&\\
& \tenscr{P}_{l}^-\left(\zeta_l\right)
\end{array}\right), \quad\text{and}\quad 
\tenscr{P}_{l}^+\left(\zeta_l\right) = {\tenscr{P}_{l}^-\left(\zeta_l\right)}^{-1} =
\left(\begin{array}{cc}
\e^{\imunit \kappa_{l;1} \zeta_l} \\ & \e^{\imunit \kappa_{l;2} \zeta_l}
\end{array}\right). \quad\text{Thus,}\label{eq:PplmiDef}\\
\vect{A}_l &=& \tens{W}_l^{-1}(0) \tens{W}_{l-1}(d_{l-1}) \vect{A}_{l-1} = \tens{W}_l^{-1}(0) \tens{W}_{l-1}(0) \tenscr{P}_{l-1}(d_{l-1}) \vect{A}_{l-1}  = \tens{Z}_{l-1} \vect{A}_{l-1}, \label{eq:ModusAmplTrMatr}
\end{eqnarray}
\end{widetext}
where we introduced the \textit{(mode) amplitude transfer matrix} \(\tens{Z}_l\).
It can be seen that the \(4\times4\) matrix \(\tens{W}_l(0)\) and its inverse consist of \(2\times2\) blocks have a special symmetry
\begin{equation}
\tens{W}_l(0) = \left(\begin{array}{cc} \tens{U}_l & \tens{U}_l \\ \imunit  \tens{\kappa}_l \tens{U}_l & -\imunit  \tens{\kappa}_l \tens{U}_l \end{array}\right),
\quad\text{and} \label{eq:Wmatr}
\end{equation}
\begin{equation}
\tens{W}_l(0)^{-1} = \frac{1}{2}\left(\begin{array}{cc} \tens{U}_l^{-1} & -\imunit  \tens{U}_l^{-1}\tens{\kappa}_l^{-1}  \\ \tens{U}_l^{-1}  & \imunit  \tens{U}_l^{-1}\tens{\kappa}_l^{-1}\end{array}\right),
\label{eq:WmatrInv0}
\end{equation}
where \(\tens{U}_l = \left(\vect{u}_{l;1}\quad \vect{u}_{l;2}\right)\) is a unitary matrix composed of the eigen vectors \(\vect{u}_{l},\) and
\begin{equation}
\tens{\kappa}_l =  \tens{U}_l \left( \begin{array}{cc} \kappa_{l;1}\\ & \kappa_{l;2} \end{array}\right) \tens{U}_l^{-1}.\label{eq:kappaMatrDef}
\end{equation}

To clarify the relation to the characteristic matrix \(\tens{L}\) defined in (\ref{eq:LDeak}), we may introduce the \textit{wave transfer matrix}
\begin{eqnarray}
\!&\!\tens{\irott{W}}_l(d_l) = \tens{W}_{l+1}(0)\tens{Z}_{l}(d_l)\tens{W}_{l}^{-1}(0) = \tens{W}_{l}(0)\tenscr{P}_{l}(d_l)\tens{W}_{l}^{-1}(0),\nonumber\\
\!&\!\text{\negthickspace{}where}\quad\left(\!\begin{array}{c}\psi_{l}(z_l+d_l) \\ \partial_z \psi_{l}(z_l+d_l)\end{array}\!\right) = \tens{\irott{W}}_{l}(d_{l})
\left(\!\begin{array}{c}\psi_{l}(z_l) \\ \partial_z \psi_{l}(z_l)\end{array}\!\right).\label{eq:WaveFuncTransfMatr}
\end{eqnarray}
Comparing equations (\ref{eq:WaveFuncTransfMatr}) and (\ref{eq:NNeuform}), and remembering that \(\Phi = \partial_z \Psi /(\imunit k \sin\theta),\) we get the following relations between the \(2\times 2\) submatrices
\begin{eqnarray}
\tens{\irott{W}}_{[11]} &=& \tens{L}_{[22]},\quad \tens{\irott{W}}_{[12]} = \frac{1}{\imunit k\sin\theta} \tens{L}_{[21]},\nonumber \\
\tens{\irott{W}}_{[21]} &=& \imunit k\sin\theta \tens{L}_{[12]}, \quad \tens{\irott{W}}_{[22]} = \tens{L}_{[11]}. \label{eq:DeakLEsWmatr}
\end{eqnarray}
Similarly, it can be shown that \(\tens{\gamma_l} = (k \sin \theta)^{-1} \tens{\kappa}_l.\)

\subsection{Parratt formalism for anisotropic layers}
To derive the generalized Parratt method, we start from the expression (\ref{eq:ModusAmplTrMatr}), which can be written in the form
\begin{equation}
\left(\begin{array}{c}\irottv{A}^{+}_{l+1} \\ \irottv{A}^{-}_{l+1}\end{array}\right) = \tens{Z}_{l} \left(\begin{array}{c}\irottv{A}^{+}_{l} \\ \irottv{A}^{-}_{l}\end{array}\right) = \left(\begin{array}{cc}^l\tens{Z}_{[11]}& ^l\tens{Z}_{[12]}\\^l\tens{Z}_{[21]}&^l\tens{Z}_{[22]}\end{array}\right) \left(\begin{array}{c}\irottv{A}^{+}_{l} \\ \irottv{A}^{-}_{l}\end{array}\right).
\end{equation}
Let us assume that \(\irottv{A}^{-}_{l} = \tenscr{R}_{l,l+1} \irottv{A}^{+}_{l} = \tenscr{R}_l \irottv{A}^{+}_{l},\) and where \(\tenscr{R}_l\) gives the \textit{amplitude reflection} for a fictional sample, where we removed the layers situated above the \(l+1\)-th layer, and the incident beam arrives from a medium which has the same material as the \(l\)-th layer. Substituting the \(\irottv{A}^{-}\) terms in the former equation, we get
\begin{equation}
\left(\begin{array}{c}\irottv{A}^{+}_{l+1} \\ \tenscr{R}_{l+1}\irottv{A}^{+}_{l+1}\end{array}\right) = \tens{Z}_{l} \left(\begin{array}{c}\irottv{A}^{+}_{l} \\ \tenscr{R}_{l}\irottv{A}^{+}_{l}\end{array}\right)
. \label{eq:PArrattFirst}
\end{equation}

From this, we can derive a recursive expression for \(\tenscr{R}_{l}:\)
\begin{equation}
\begin{split}
\tenscr{R}_{l} &=
\left({^l\tens{Z}_{[22]}} - \tenscr{R}_{l+1} {^l\tens{Z}_{[12]}} \right)^{-1}
\left(\tenscr{R}_{l+1} {^l\tens{Z}_{[11]}} - {^l\tens{Z}_{[21]}}\right)\\
&=
\left(\tens{I} - {^l\tens{Z}_{[22]}}^{-1} \tenscr{R}_{l+1} {^l\tens{Z}_{[12]}} \right)^{-1}\\
&\left({^l\tens{Z}_{[22]}}^{-1} \tenscr{R}_{l+1} {^l\tens{Z}_{[11]}} - {^l\tens{Z}_{[22]}}^{-1} {^l\tens{Z}_{[21]}} \right).
\end{split}\label{eq:ParrattWithZ}
\end{equation}
This is not what we usually find in papers.
The \textit{`wave' reflection matrix} \(\tensfr{R}_l,\) connecting the backwards \(\left(\psi^-_l\right)\) and forwards \(\left(\psi^+_l\right)\) propagating wave components of \(\psi_l = \psi^-_l + \psi^+_l\) by \(\psi^-_l(\zeta_l) = \tensfr{R}_l(\zeta_l)\psi^+_l(\zeta_l),\)
will be
\begin{equation}
\tensfr{R}_l(\zeta_l) = \tens{U}_l \tenscr{P}_l^-(\zeta_l) \tenscr{R}_l \tenscr{P}_l^-(\zeta_l) \tens{U}_l^{-1},\label{eq:waveReflAmplRefl}
\end{equation}
as \(\psi^\pm_l(\zeta_l) = \tens{U}_l\tenscr{P}^\pm_l(\zeta_l)\irottv{A}^\pm_l\) according to (\ref{eq:psilAlak}).
After some lengthy algebraic transformations (see Appendix \ref{app:DerivParratt}), using the notation
\begin{equation}
\tenshat{R}_l = \tensfr{R}_l(0) = \tens{U}_l \tenscr{R}_l \tens{U}_l^{-1},\label{eq:ReflHatAmplRefl}
\end{equation}
the generalized Parratt formula can be written as
\begin{eqnarray}
\tenshat{R}_{l} &= \tens{P}_l^+
\left(
\tens{I} +
\tenscr{D}_{l+1,l}\tenshat{R}_{l+1}\tenscr{D}_{l+1,l}^{-1} \tenscr{F}_{l,l+1}
\right)^{-1} \nonumber \\
&\left(
\tenscr{F}_{l,l+1} +
\tenscr{D}_{l+1,l} \tenshat{R}_{l+1}\tenscr{D}_{l+1,l}^{-1}
\right)
\tens{P}_l^+, \label{eq:ParrattGenForm}
\end{eqnarray}
where the Fresnel reflection \(\tenscr{F}\) and transmission \(\tenscr{D}\) matrices belonging to a single interface situated between the \(l\)-th and \(l+1\)-th layers are
\begin{eqnarray}
\tenscr{F}_{l,l+1} &=& \left[ \tens{\kappa}_{l} + \tens{\kappa}_{l+1} \right]^{-1}
 \left[ \tens{\kappa}_{l} - \tens{\kappa}_{l+1} \right], \nonumber\\
 \tenscr{D}_{l,l+1} &=& 2 \left[\tens{\kappa}_{l} + \tens{\kappa}_{l+1} \right]^{-1} \tens{\kappa}_{l},\label{eq:FresnelRT}
\end{eqnarray}
and we introduced the notation
\begin{equation}
\tens{P}_l^+ = \tens{U}_{l} \tenscr{P}_{l}^+(d_l) \tens{U}_{l}^{-1}.\label{eq:PeqUPU}
\end{equation}

The \(\tenscr{D}_{l+1,l}\)  matrix corresponds to the transmission in the opposite direction through the \(l, l+1\) interface, which is related to the reflected wave propagating in the direction away from the substrate towards the sample surface.
In the isotropic case, our matrices will be scalars, and hence
\begin{equation}
R_{l} = {P_l^+}^2 \left( 1 + R_{l+1} F_{l,l+1} \right)^{-1}  \left(F_{l,l+1} + R_{l+1} \right), \label{eq:ParrattIsotrop}
\end{equation}
which is in good agreement with the expression given by Parratt in article \cite{Parratt-1954}, i.e., eq.~(\ref{eq:Parratt}). Here, the term \({P_l^+}^2\) corresponds to the term of \(a_{n-1}^4\) of \cite{Parratt-1954}. We have a power of 2 instead of the power of 4 in (\ref{eq:ParrattIsotrop}), as we have not assumed our amplitudes in the middle of the layers. Thus, \(P_l^+\) corresponds to the whole layer thickness, and not only to a half layer thickness as \(a_{n-1}\) is.

Following the assumptions of Parratt mentioned in section \ref{sec:OrigParrattForm}, we can also assume that \(\tenshat{R}_{\Omega+1} = \tenshat{R}_{\Omega+1,\Omega+2} = 0,\) thus
\begin{equation}
\tenshat{R}_{\Omega} = \tens{P}_{\Omega}^+ \tenscr{F}_{\Omega,\Omega+1} \tens{P}_{\Omega}^+. \label{eq:ROmega}
\end{equation}
This is the reflection matrix belonging to the last effectively reflecting interface. The reflectivity matrix for the whole sample, which we measure using reflectometry, will be \(\tenshat{R}_{0,1},\) and \(\tens{P}_0^+ = \tens{I},\) as the \(0\)-th layer is the medium above the sample surface, and that is usually air, in which we neglect the absorption along the source-sample and sample-detector paths.

In the expression (\ref{eq:ParrattGenForm}), we have exponential terms, which cause the numerical instability for `very thick' samples,  only in \(\tens{P}_l^+\) matrices. But here, we have only exponentially decreasing terms. Therefore, this method should be devoid of such instability. We may have problems with matrix inversions in the case of ill-conditioned matrices of equations (\ref{eq:ParrattGenForm}) and (\ref{eq:FresnelRT}).

In transfer matrix methods, the exponentially decreasing \(\tenscr{P}_l^+\) and increasing \(\tenscr{P}_l^-\) terms always appear together, which is why they may become unstable. This is obvious in eq.~(\ref{eq:LDeak}), as both are represented in the \(\sinh\) and \(\cosh\) functions.

It is easy to see that if we express \(\tens{R}_l = \tensfr{R}_l(\zeta_l = d_l) = \tens{P}_l^- \tenshat{R}_l \tens{P}_l^-\) instead of \(\tenshat{R}_l = \tensfr{R}_l(\zeta_l = 0),\) we get a slightly different form of the generalized Parratt formula
\begin{eqnarray}
\tens{R}_{l} &=
\left(
\tens{I} +
\tenscr{D}_{l+1,l}\tenshat{R}_{l+1}\tenscr{D}_{l+1,l}^{-1} \tenscr{F}_{l,l+1}
\right)^{-1} \nonumber \\
&\left(
\tenscr{F}_{l,l+1} +
\tenscr{D}_{l+1,l} \tenshat{R}_{l+1}\tenscr{D}_{l+1,l}^{-1}
\right), \label{eq:ParrattGenFormB}
\end{eqnarray}
which we have already obtained in (\ref{deakParratt}). (We need to recognize that \(\exp\left(\imunit kd_{l}\tens{\gamma}_{l}\sin\theta\right) = \tens{P}^+_l,\) \(\tens{r}_l = \tenscr{F}_{l-1,l}\) and \(\tens{I} - \tens{r}_l = \tenscr{D}_{l,l-1}.\))
It is clear from definition of \(\tensfr{R}_l,\) that \(\tens{R}_l\) is the reflection matrix which connects the backward and forward propagating waves just above the interface of the \(l\)-th and \(l+1\)-th layer; and \(\tenshat{R}_l\) connects the waves just below the interface of the \(l-1\)-th and \(l\)-th layer. This the reason \(\tenshat{R}_l = \tens{P}^+_l \tens{R}_l \tens{P}^+_l,\) as we need to propagate the waves from \(z_l\) to \(z_{l-1}.\) As we have seen already, at the surface of the sample, there will be no difference between the two approaches, as \(\tens{P^+_0} =\tens{I}\) practically. It is easier to discover the similarities between (\ref{eq:ParrattGenForm}) and the original Parratt formula (\ref{eq:Parratt}) than between (\ref{eq:ParrattGenFormB}) and (\ref{eq:Parratt}).

\subsubsection{Transmission}
For the calculation of transmission, we use a different approach. Let denote by \(\tenscr{T}_{l}\) and \(\tens{T}_{l}\) the matrices for the \(l\)-th layer defined by
\begin{equation}
\irottv{A}^{+}_{l+1} = \tenscr{T}_{l}\irottv{A}^{+}_{l}, \quad \tens{T}_{l} = \tens{U}_{l+1} \tenscr{T}_{l} \tens{U}_{l}^{-1}, \label{eq:ParrattTDef}
\end{equation}
where these matrices will represent the transmission through the bulk of the \(l\)-th layer and the \(l,l+1\)-th interface for amplitudes and wave functions, respectively.
The transmission matrix for the whole sample will be the product of such matrices
\begin{equation}
\tens{T} = \tens{T}_{\Omega} \cdots \tens{T}_2\tens{T}_1\tens{T}_0.
\end{equation}
Using (\ref{eq:PArrattFirst}), (\ref{eq:ReflHatAmplRefl}), and (\ref{eq:ParrattTDef}), we may write
\begin{eqnarray}
\tenscr{T}_{l} &=& {^l\tens{Z}_{[11]}} + {^l\tens{Z}_{[12]}} \tenscr{R}_{l},\quad\text{and} \nonumber\\
\tens{T}_{l} &=& \tens{U}_{l+1}{^l\tens{Z}_{[11]}} \tens{U}_{l}^{-1} + \tens{U}_{l+1} {^l\tens{Z}_{[12]}}  \tens{U}_{l}^{-1} \tenshat{R}_{l},\label{eq:TransmissionStart}
\end{eqnarray}
from which, after some algebraic transformations (see Appendix \ref{app:DerivParratt}), we obtain
\begin{eqnarray}
\tens{T}_{l} &= \tenscr{D}_{l+1,l}^{-1} \left[
\tens{I} - \tenscr{F}_{l,l+1}
\left(
\tens{I} +
\tenscr{D}_{l+1,l}\tenshat{R}_{l+1}\tenscr{D}_{l+1,l}^{-1} \tenscr{F}_{l,l+1}
\right)^{-1}\right.\nonumber\\
&\left.\left(
\tenscr{F}_{l,l+1} +
\tenscr{D}_{l+1,l} \tenshat{R}_{l+1}\tenscr{D}_{l+1,l}^{-1}
\right)
\right] \tens{P}_l^+. \label{eq:TransmissIteration}
\end{eqnarray}
This formula can be regarded as numerically stable, given that \(\tens{P}_l^+\) includes only exponentially decreasing terms and that the stability of Eq.~(\ref{eq:ParrattGenForm}) used to calculate \(\tenshat{R}_{l+1}\) has been shown. Using \(\tenshat{R}_{\Omega+1} = \tenshat{R}_{\Omega+1,\Omega+2} = 0\) again and some algebra, we obtain
\begin{equation}
\tens{T}_{\Omega} = \tenscr{D}_{\Omega+1,\Omega}^{-1}\left(\tens{I} - \tenscr{F}_{\Omega,\Omega+1}^2\right) \tens{P}_\Omega^+ = \tenscr{D}_{\Omega,\Omega+1} \tens{P}_\Omega^+.
\end{equation}
In the case of a single interface, taking \(\tens{P}^+_0 = \tens{I}\) simplifies the result to the expected Fresnel formula, \(\tens{T}_{0} = \tenscr{D}_{0,1}.\)

\subsection{Roughness}
We rarely have sharp interfaces in practice, hence any useful method should also take into account interface roughness. There are several approaches to tackle this problem.
The `brute force method' models the interface by dividing it into many thin layers, effectively approximating the susceptibility, scattering length density, and magnetization profiles across the interface as a step function. This works well, but it significantly slows the calculations. Therefore, analytical approximations are preferred whenever possible. Here we provide two such approximations.

\subsubsection{First analytical roughness approximation}\label{ssubsec:Tolan}
First, we follow the train of thought available in \cite{Daillant-b2009-chp3, TolanPhd}. If the rough interface of the \(l\)-th and \(l+1\)-th layer is at depth \(z_{l+1} + \Delta z_{l+1}(x,y),\) then the corrected \(\tens{Z}\) matrix taking into account roughness on this interface is
\begin{equation}
\begin{split}
\tens{Z}^\text{rough}_{l} &= \left<\tenscr{P}_{l+1}^{-1}(\Delta z_{l+1}(x,y)) \tens{W}_{l+1}^{-1}(0) \tens{W}_{l}(0) \right.\\ &\left.\tenscr{P}_{l}(\Delta z_{l+1}(x,y)) \tenscr{P}_{l}(d_l)\right>, \label{eq:ZRoughStartAver}
\end{split}
\end{equation}
where the angled brackets denote the averaging along the interface, i.e., the \(x\) and \(y\) coordinates. The terms \(\tenscr{P}_{l}(\Delta z_{l+1}(x,y)),\) \(\tenscr{P}_{l+1}^{-1}(\Delta z_{l+1}(x,y))\) correspond to the `phase shifts' arising from the locale shift
\(\Delta z\) of the interface.
Assuming that the interface is a result of an ergodic random process and the height probability distribution is Gaussian with standard deviation \(\sigma_{l} = \sigma_{l,l+1},\) which is the rms roughness of the interface in depth \(z_{l+1},\) we will have
\begin{equation}
\tens{Z}^\text{rough}_{l} = \tens{G}_l \odot \left(\tens{W}_{l+1}^{-1}(0) \tens{W}_{l}(0)\right) \tenscr{P}_{l}(d_l), \label{eq:ZRoughCorr}
\end{equation}
where \(\odot\) denotes the Hadamard product \(( (\tens{A} \odot \tens{B})_{ij}  = A_{ij} B_{ij} )\) or element-wise product, and
\begin{eqnarray}
\tens{G}_l \!&\!=\!&\! \left(\begin{array}{cc}\tens{G}_l^- & \tens{G}_l^+ \\ \tens{G}_l^+ & \tens{G}_l^-\end{array}\right),\nonumber \\
\tens{G}_l^\pm \!&\!=\!&\! \left(\begin{array}{cc}
\e^{-\frac{1}{2}\sigma_l^2\left(\kappa_{l;1}\pm\kappa_{l+1;1}\right)^2} & \e^{-\frac{1}{2}\sigma_l^2\left(\kappa_{l;2}\pm\kappa_{l+1;1}\right)^2} \\
\e^{-\frac{1}{2}\sigma_l^2\left(\kappa_{l;1}\pm\kappa_{l+1;2}\right)^2} & \e^{-\frac{1}{2}\sigma_l^2\left(\kappa_{l;2}\pm\kappa_{l+1;2}\right)^2}\end{array}\right). \label{eq:GDef}
\end{eqnarray}

The equation (\ref{eq:ZRoughCorr}) together with (\ref{eq:GDef}) is structurally very similar to the result of Ref.~\cite{Vidal-1984} derived for isotropic systems.
After some algebra we obtained (see Appendix \ref{app:DerivRoughParratt}) the formulae
\begin{equation}
\begin{split}
\tenshat{R}_{l}&=
\tens{P}_l^+ \left(\tens{I} - {\tens{\delta}^+_l}^{-1} \tenshat{R}_{l+1} {\tens{\delta}^-_l} \right)^{-1} \\
 &\left( {\tens{\delta}^+_l}^{-1} \tenshat{R}_{l+1} {\tens{\delta}^+_l}  -  {\tens{\delta}^+_l}^{-1} {\tens{\delta}^-_l} \right)
 \tens{P}_l^+,\\
\tens{T}_{l}\!&\!= \left[ \tens{\delta}^+_l + \tens{\delta}^-_l
\left(\tens{I} - {\tens{\delta}^+_l}^{-1} \tenshat{R}_{l+1} {\tens{\delta}^-_l} \right)^{-1} \right.\\
 &\left.\left( {\tens{\delta}^+_l}^{-1} \tenshat{R}_{l+1} {\tens{\delta}^+_l}  -  {\tens{\delta}^+_l}^{-1} {\tens{\delta}^-_l} \right)
\right] \tens{P}_l^+,
 \quad \text{where} \\
 \tens{\delta}^+_l &= \tens{U}_{l+1} \left[\tens{G}_l^- \odot \left(\tens{U}_{l+1}^{-1} \tenscr{D}_{l+1,l}^{-1} \tens{U}_{l}\right)\right] \tens{U}_{l}^{-1},\\
  \tens{\delta}^-_l &= -\tens{U}_{l+1} \left[\tens{G}_l^+ \odot \left(\tens{U}_{l+1}^{-1} \tenscr{D}_{l+1,l}^{-1}\tenscr{F}_{l,l+1} \tens{U}_{l}\right)\right] \tens{U}_{l}^{-1}.
\end{split}\label{eq:RoughParrattGenForm}
\end{equation}
It is straightforward to show that when the roughness \(\sigma_{l}\) is 0, every element of the matrix \(\tens{G}\) equals 1. Such matrices act as the identity under the Hadamard product, and consequently, the formulas (\ref{eq:ParrattGenForm}) and (\ref{eq:TransmissIteration}) derived for systems without roughness are recovered, as expected.

In the isotropic case, our matrices will be scalars and therefore
\begin{eqnarray}
R_{l} &= {P_l^+}^2 \left( 1 + R_{l+1} F_{l,l+1} \e^{-2\sigma^2_l\kappa_{l}\kappa_{l+1} } \right)^{-1} \nonumber\\ &\left(F_{l,l+1}\e^{-2\sigma^2_l\kappa_{l}\kappa_{l+1} }  + R_{l+1} \right),
\end{eqnarray}
where we have the Névot\,--\,Croce factor \(\e^{-2\sigma^2_l\kappa_{l}\kappa_{l+1} }\) \cite{NevotCroce-1980, Vidal-1984, RohlsbergerHypInt99, Esashi-2021}. An approximation of this is provided by the Debye\,--\,Waller factor \(\approx\e^{-\frac{1}{2}\sigma^2_l q_z^2 },\) where \(\vect{q}\) is the impulse transfer, and we apply the \(\kappa_{l} \approx \kappa_{l+1} \approx \frac{1}{2}q_z\) approximations. Both of these factors appear frequently in the literature \cite{Daillant-b2009-chp3, Basu-b2022-chp2, RohlsbergerHypInt99, Vidal-1984, Esashi-2021}.

Following the assumptions of Parratt, we can also assume that \(\tenshat{R}_{\Omega+1} = \tenshat{R}_{\Omega+1,\Omega+2} = 0,\) and thus, the initial value of the recursion will be
\begin{equation*}
\begin{split}
&\tenshat{R}_{\Omega}=
\tens{P}_{\Omega}^+
 \left(  -  {\tens{\delta}^+_{\Omega}}^{-1} {\tens{\delta}^-_{\Omega}} \right)
 \tens{P}_{\Omega}^+
 =
\tens{P}_{\Omega}^+
\tens{U}_\Omega
\left[\tens{G}_\Omega^- \odot \left(\tens{U}_{\Omega+1}^{-1} \tenscr{D}_{\Omega+1,\Omega}^{-1} \right.\right.
\\ &\left.\left.
\tens{U}_{\Omega}\right)\right]^{-1}
\left[\tens{G}_\Omega^+ \odot \left(\tens{U}_{\Omega+1}^{-1} \tenscr{D}_{\Omega+1,\Omega}^{-1}\tenscr{F}_{\Omega,\Omega+1} \tens{U}_\Omega\right)\right]
\tens{U}_\Omega^{-1}
\tens{P}_{\Omega}^+.
\end{split}
\end{equation*}

\subsubsection{Second analytical roughness approximation}
According to Röhlsberger \cite{RohlsbergerHypInt99,Rohlsberger-b2009-chp4}, a rough interface can be taken into account by replacing the products of the transfer matrices with a third matrix inserted between them. The new matrix takes into account the roughness. This is derived using the Campbell\,--\,Baker\,--\,Hausdorff formula. For further details, please see the original works. We obtained the following results (Appendix \ref{app:Rohlsberger}). The roughness can be incorporated by applying the generalized Parratt formulae (\ref{eq:ParrattGenForm}) and (\ref{eq:TransmissIteration}), substituting the Fresnel-coefficient matrices with
\begin{eqnarray}
^{R}\tenscr{F}_{l,l+1} \negthickspace &=&\negthickspace \left[\tens{\eta}_{l,l+1}^{-1} \tens{\kappa}_{l} + \tens{\kappa}_{l+1} \tens{\eta}_{l,l+1}\right]^{-1}\negthickspace
 \left[\tens{\eta}_{l,l+1}^{-1} \tens{\kappa}_{l} - \tens{\kappa}_{l+1}\tens{\eta}_{l,l+1} \right]\negthickspace ,\nonumber\\
^{R}\tenscr{D}_{l,l+1} \negthickspace&=&\negthickspace 2 \left[\tens{\eta}_{l,l+1}^{-1}\tens{\kappa}_{l} + \tens{\kappa}_{l+1} \tens{\eta}_{l,l+1} \right]^{-1}\negthickspace \tens{\kappa}_{l},\label{eq:FresnelRTRough}
\end{eqnarray}
where
\begin{eqnarray}
\tens{\eta}_{l,l+1} &= \exp\left[\sum\limits_{n=1}^\infty \frac{1}{n!} \left(\frac{\sigma_{l,l+1}^2}{2}\right)^n \tens{c}_n\right], \quad
\tens{c}_1 = \tens{\kappa}_{l+1}^2 - \tens{\kappa}_l^2,\nonumber\\ &\tens{c}_{n+1} = - \left[\tens{c}_n \left(2\tens{\kappa}_{l+1}^2 + \tens{\kappa}_l^2\right) + \tens{\kappa}_l^2 \tens{c}_n\right].
\end{eqnarray}
If the roughness \(\sigma\) is 0, then \(\tens{\eta}_{l,l+1}\) will be the identity matrix, and we will recover the Fresnel-coefficient matrices given in eq.~(\ref{eq:FresnelRT}).

The principal conceptual difference between the two approximations can be summarized as follows: Here, we calculate the reflection and transmission matrices assuming a \(\left<\tens{\chi}\right>(z)\) depth profile averaged laterally, thus depending only on \(z.\) In the previous approximation of \ref{ssubsec:Tolan}, we average the matrices representing an ensemble of intermediate bilayers, where the position of the boundary of these bilayers is changing according to the distribution characterizing the rough interface. The seminal articles \cite{NevotCroce-1980} and \cite{Vidal-1984} are also based on the latter concept used in \ref{ssubsec:Tolan}.

\section{Examples, comparison with other methods\label{sec:ExamplesCompar}}

To compare the methods and verify the generalized Parratt formalism presented here, we performed calculations for different multilayer structures. For the calculations, we used the program FitSuite (version 2.3) \cite{Sajti2009}, developed by our group over the past two decades. We should mention that its precursor, the program EFFI \cite{Spiering00}, from which most of the calculation routines are derived, has an even longer history. Accordingly, the program is relatively well-tested in terms of reflectometry. Thus, it
is reasonable to test our new generalized Parratt method presented here and implemented in FitSuite, using the previously implemented, tested methods based on \cite{Deak96, Deak01, DeakPhd}. All the program codes used for calculation are attached with their documentation to the FitSuite program, which is freely downloadable.

\begin{figure*}[htb]
\includegraphics[width=7.5cm, clip]{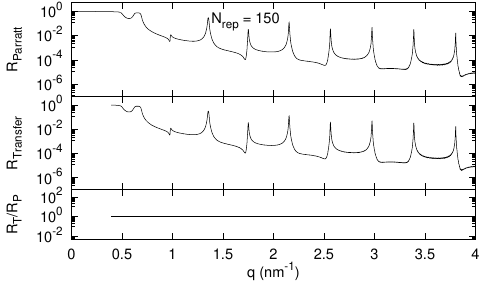}~
\includegraphics[width=7.5cm, clip]{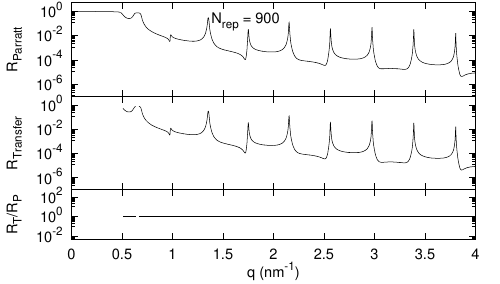}
\caption{Comparison of X-ray reflectometry results for [Ni(7 nm)/Ti(8 nm)] \(_{N_\text{rep}}\)/Glass multilayer systems for different bilayer repetition numbers \(N_\text{rep}.\) The reflectivity was computed using both the  transfer matrix method \(\left(R_\text{Transfer}\right)\) and the generalized Parratt algorithm \(\left(R_\text{Parratt}\right)\) described here. The ratio of these results \(\left(R_\text{T}/R_\text{P} = R_\text{Transfer}/R_\text{Parratt}\right)\) is also plotted. The results of numerical instabilities are apparent in the total reflection region and near the first Bragg-peak for \(N_\text{rep} = 900.\)}\label{fig:NiTiStruct}
\end{figure*}
We plotted in Fig.~\ref{fig:NiTiStruct} the results of X-ray reflectometry (\(\lambda =\) 0.154 nm) calculations for [Ni(7 nm)/Ti(8 nm)]\(_{N_\text{rep}}\)/Glass multilayers, where \(N_\text{rep}\) is the repetition number of the Ni/Ti bilayers. The roughness is assumed to be 0. We found instability in the total reflection region using the transfer matrix method; this is why the curves do not start at \(q=0.\) We also observed instability at the first Bragg peak with this method, which produced \textit{NaN (Not a Number)} values for \(N_\text{rep} = 900.\)
The Parratt algorithm, as we expected, had no such problems. For greater grazing angles, the two methods give the same results.

\begin{figure*}[htb]
\includegraphics[width=7.5cm, clip]{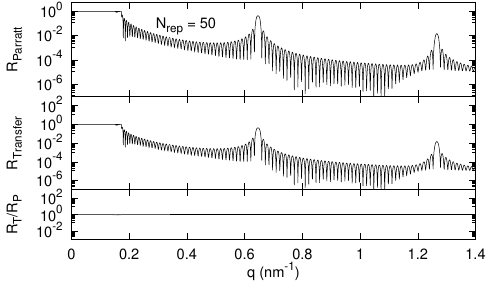}~
\includegraphics[width=7.5cm, clip]{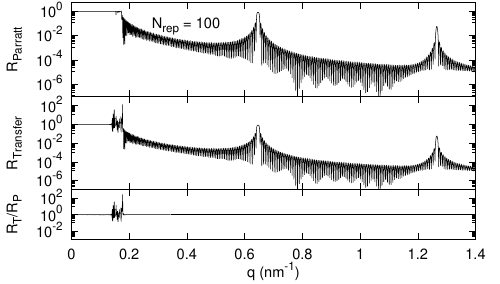}
\includegraphics[width=7.5cm, clip]{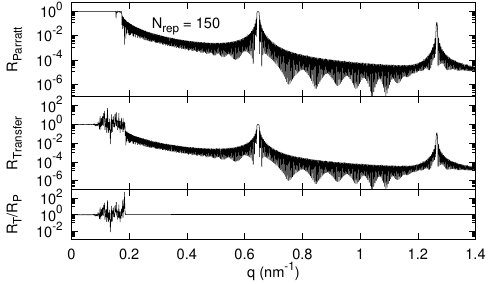}~
\includegraphics[width=7.5cm, clip]{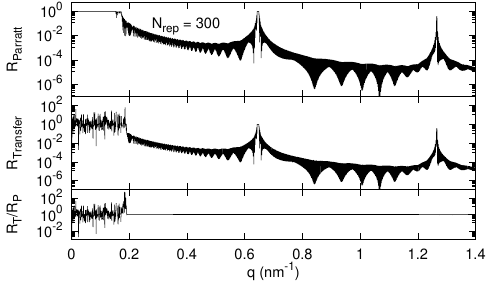}
\includegraphics[width=7.5cm, clip]{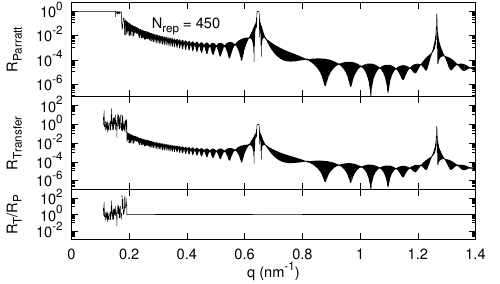}~
\includegraphics[width=7.5cm, clip]{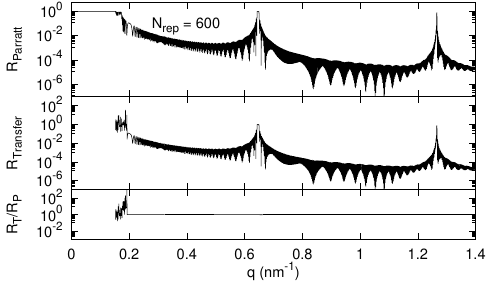}
\caption{Comparison of neutron reflectometry results for [Cr(4 nm)/Fe(6 nm)]\(_{N_\text{rep}}\)/MgO multilayer systems for different bilayer repetition numbers \(N_\text{rep}.\) The reflectivity was computed using both the transfer matrix method \(\left(R_\text{Transfer}\right)\) and the generalized Parratt algorithm \(\left(R_\text{Parratt}\right)\) described here. The ratio of these results \(\left(R_\text{T}/R_\text{P} = R_\text{Transfer}/R_\text{Parratt}\right)\) is also plotted. The results of numerical instabilities are apparent in and near the total reflection region.}\label{fig:FeCrStruct}
\end{figure*}
We plotted in Fig.~\ref{fig:FeCrStruct} the results of such calculations for [Cr(4 nm)/Fe(6 nm)]\(_{N_\text{rep}}\)/MgO multilayers, where \(N_\text{rep}\) is the repetition number of the Cr/Fe bilayers. The roughness is assumed to be 0, and the effective magnetic induction is 1 T. It is apparent that for \(N_\text{rep} = 50,\) there are no stability problems, and the two methods provide the same results. For greater repetition numbers, we have instabilities with the transfer matrix method in and near the total reflection region. For greater grazing angles, the two methods provide the same results. In the case of repetition numbers 450 and 600 for very small \(q\)-s, the transfer matrix method resulted \textit{NaN} values; this is the reason the curves do not start from \(q=0.\)

\begin{figure*}[htb]
\includegraphics[width=7.5cm, clip]{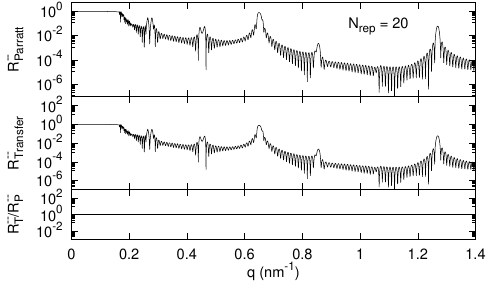}~
\includegraphics[width=7.5cm, clip]{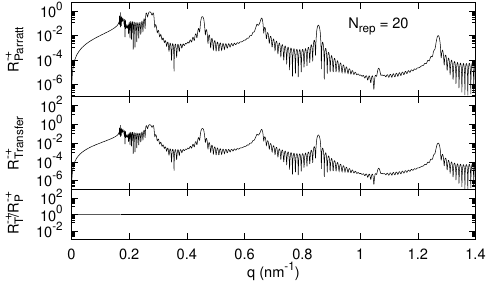}
\includegraphics[width=7.5cm, clip]{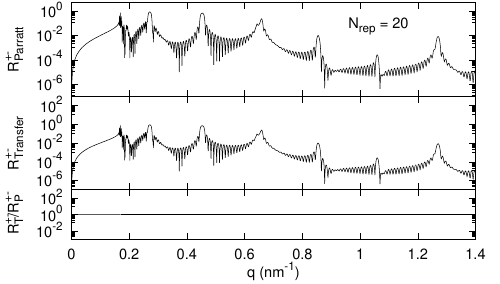}~
\includegraphics[width=7.5cm, clip]{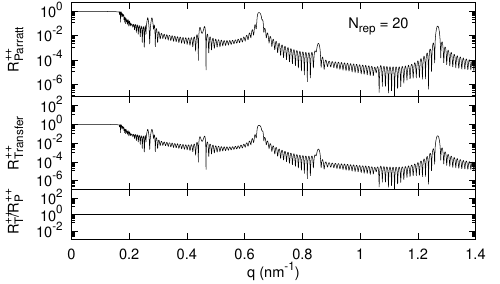}
\includegraphics[width=7.5cm, clip]{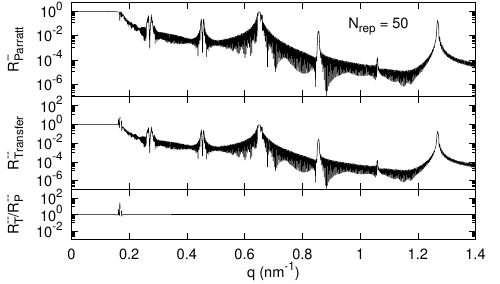}~
\includegraphics[width=7.5cm, clip]{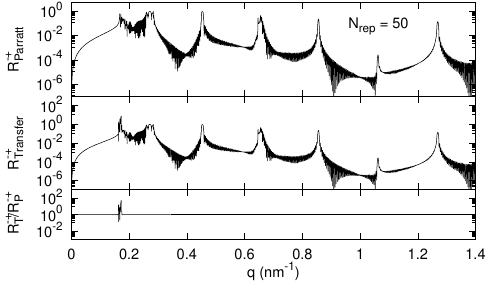}
\includegraphics[width=7.5cm, clip]{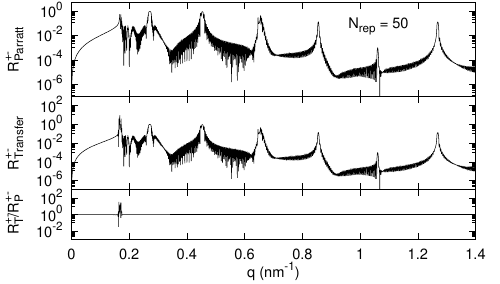}~
\includegraphics[width=7.5cm, clip]{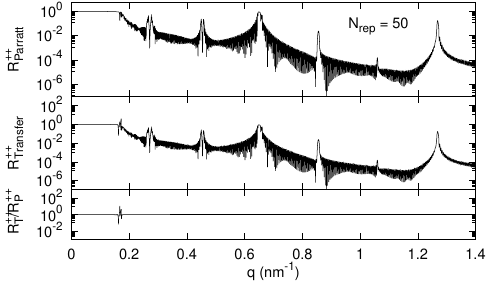}
\caption{Comparison of polarised neutron reflectometry results for [Cr(4 nm)/Fe(6 nm)/Cr(4 nm)/Fe(6 nm)/Cr(4 nm)/Fe(6 nm)]\(_{N_\text{rep}}\)/MgO multilayer systems for different hexalayer repetition numbers \(N_\text{rep}.\) In the hexalayers, the iron layers are magnetised in different directions (30°, 120°, 200° measured from the plane of incidence) in the sample plane. The reflectivities for four combined polarizer-analyzer states \((-\,-, -+, +-, ++)\) were computed using both the transfer matrix method \(\left(R_\text{Transfer}\right)\) and the generalized Parratt algorithm \(\left(R_\text{Parratt}\right)\) described here. The ratio of these results \(\left(R_\text{T}/R_\text{P} = R_\text{Transfer}/R_\text{Parratt}\right)\) is also plotted. The results of numerical instabilities are apparent in and near the total reflection region.}\label{fig:FeCrStructAniz}
\end{figure*}
\begin{figure*}[htb]
\includegraphics[width=7.5cm, clip]{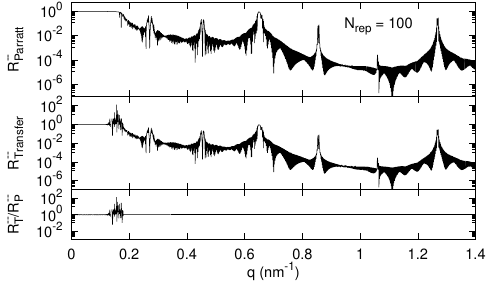}~
\includegraphics[width=7.5cm, clip]{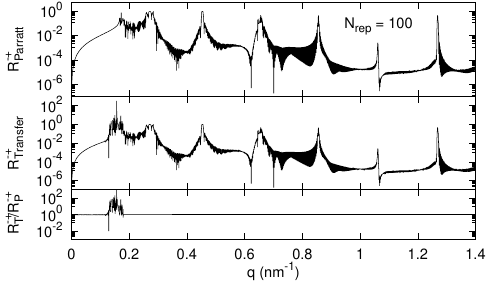}
\includegraphics[width=7.5cm, clip]{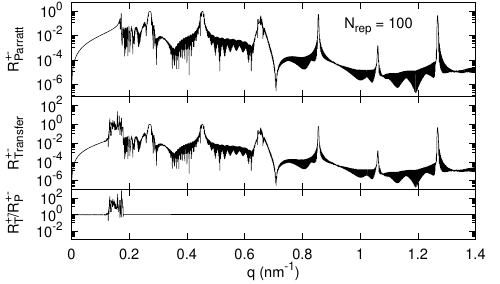}~
\includegraphics[width=7.5cm, clip]{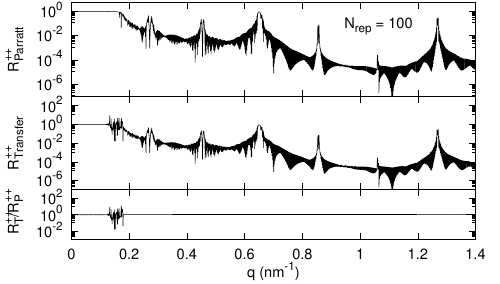}
\includegraphics[width=7.5cm, clip]{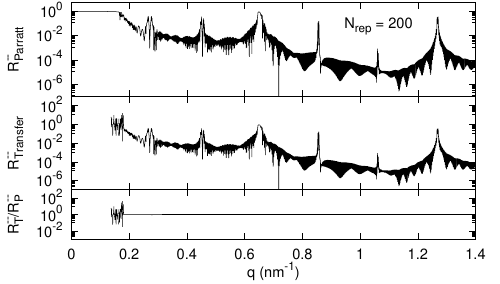}~
\includegraphics[width=7.5cm, clip]{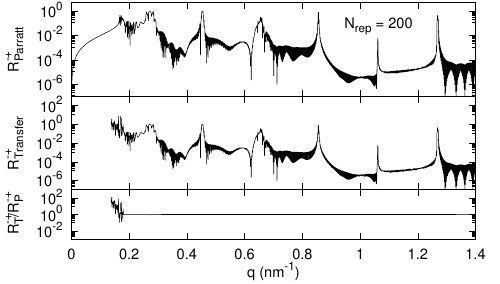}
\includegraphics[width=7.5cm, clip]{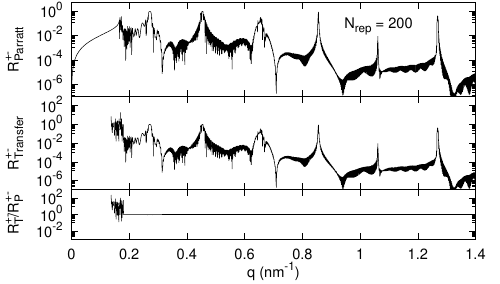}~
\includegraphics[width=7.5cm, clip]{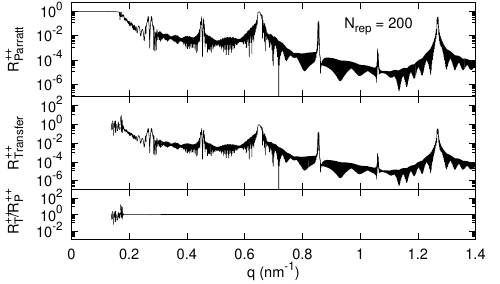}
\caption{It is the same as Fig.~\ref{fig:FeCrStructAniz}, but calculated for different repetition numbers \(N_\text{rep}.\)}\label{fig:FeCrStructAniz2}
\end{figure*}
\begin{figure*}[htb]
\includegraphics[width=7.5cm, clip]{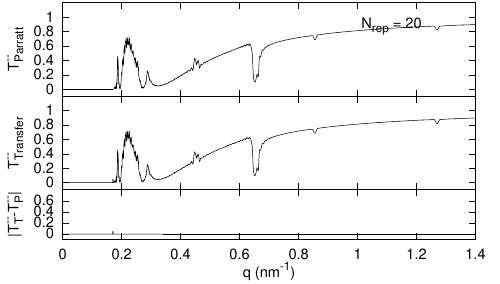}~
\includegraphics[width=7.5cm, clip]{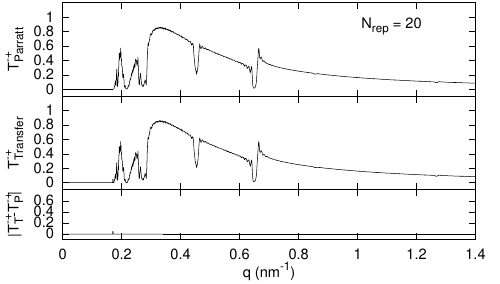}
\includegraphics[width=7.5cm, clip]{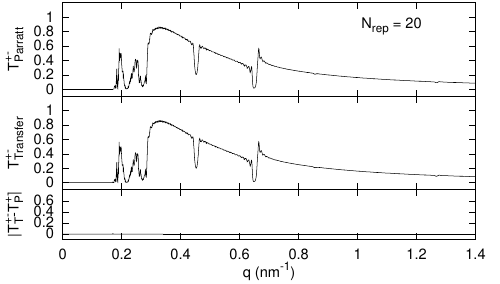}~
\includegraphics[width=7.5cm, clip]{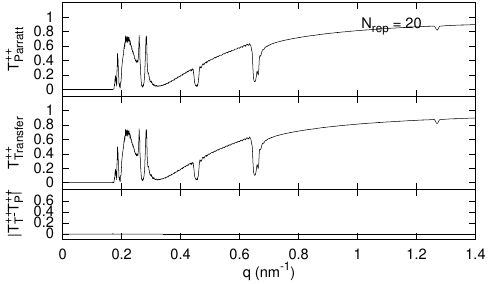}
\includegraphics[width=7.5cm, clip]{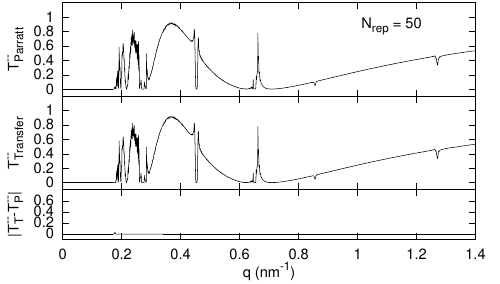}~
\includegraphics[width=7.5cm, clip]{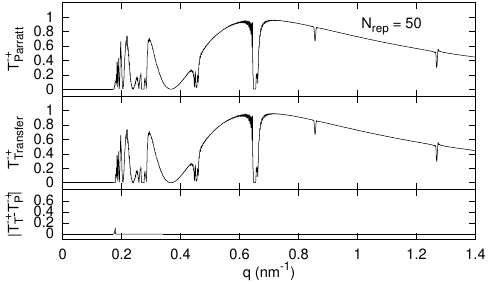}
\includegraphics[width=7.5cm, clip]{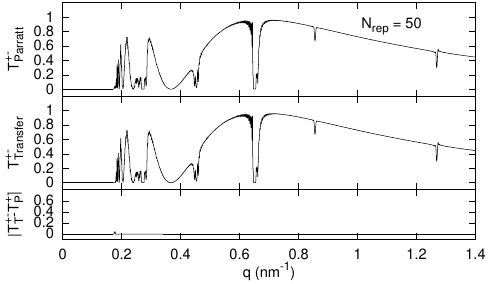}~
\includegraphics[width=7.5cm, clip]{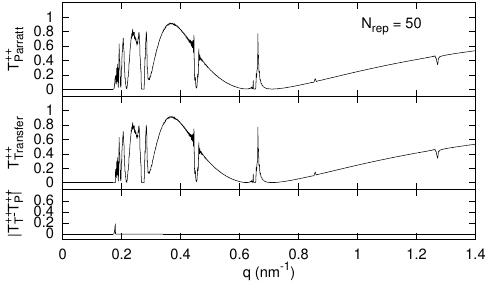}
\caption{Comparison of polarised neutron transmissivity results for [Cr(4 nm)/Fe(6 nm)/Cr(4 nm)/Fe(6 nm)/Cr(4 nm)/Fe(6 nm)]\(_{N_\text{rep}}\)/MgO(1 mm) multilayer systems for different hexalayer repetition numbers \(N_\text{rep}.\) In the hexalayers, the iron layers are magnetised in different directions (30°, 120°, 200° measured from the plane of incidence) in the sample plane. The transmissivities for four combined polarizer-analyzer states \((-\,-, -+, +-, ++)\) were computed using both the transfer matrix method \(\left(T_\text{Transfer}\right)\) and the generalized Parratt algorithm \(\left(T_\text{Parratt}\right)\) described here. The difference of these results \(\left(\left|T_\text{T}-T_\text{P}\right| = \left|T_\text{Transfer}-T_\text{Parratt}\right|\right)\) is also plotted. The results of numerical instabilities are apparent in and near the total reflection region.}\label{fig:FeCrStructAnizTransm}
\end{figure*}
\begin{figure*}[htb]
\includegraphics[width=7.5cm, clip]{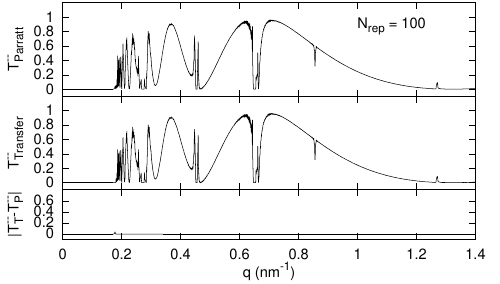}~
\includegraphics[width=7.5cm, clip]{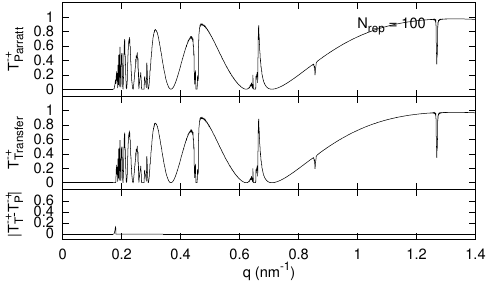}
\includegraphics[width=7.5cm, clip]{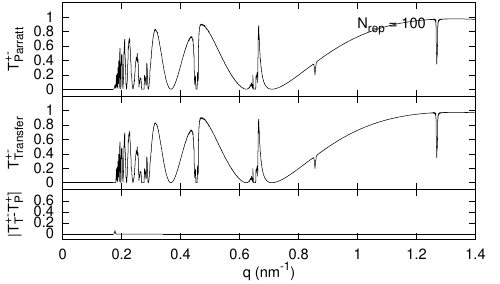}~
\includegraphics[width=7.5cm, clip]{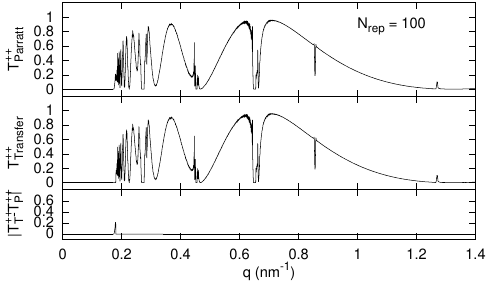}
\includegraphics[width=7.5cm, clip]{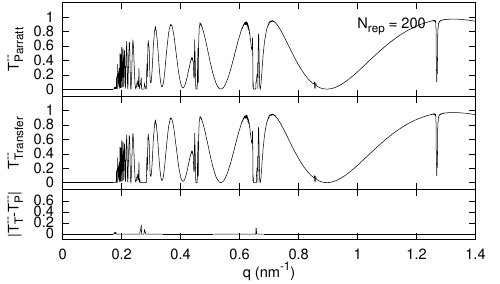}~
\includegraphics[width=7.5cm, clip]{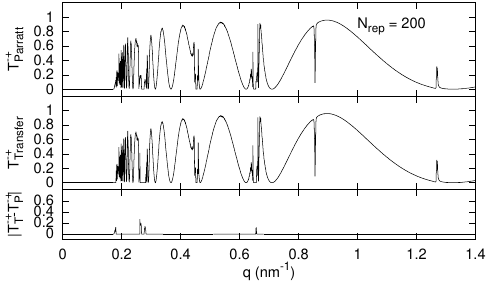}
\includegraphics[width=7.5cm, clip]{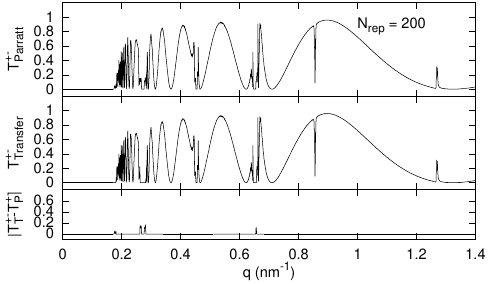}~
\includegraphics[width=7.5cm, clip]{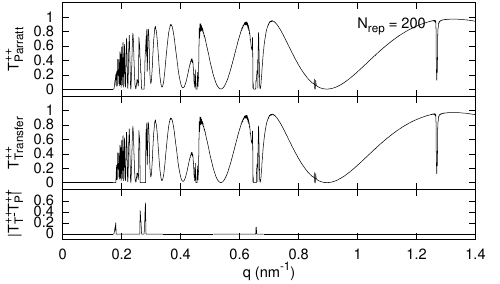}
\caption{It is the same as Fig.~\ref{fig:FeCrStructAnizTransm}, but calculated for different repetition numbers \(N_\text{rep}.\)}\label{fig:FeCrStructAnizTransm2}
\end{figure*}
To test the generalized Parratt formalism for anisotropic magnetic systems, we have chosen a [Cr(4 nm)/Fe(6 nm)/Cr(4 nm)/Fe(6 nm)/Cr(4 nm)/Fe(6 nm)]\(_{N_\text{rep}}\)/MgO(1 mm). In these hexalayers, the iron layers have different magnetization directions (30°, 120°, 200° measured from the plane of incidence) in the sample plane, with an effective magnetic induction of 1 T. The reflectivities for four combined polarizer-analyzer states \((-\,-, -+, +-, ++)\) were calculated by both methods and plotted in Figs.~\ref{fig:FeCrStructAniz}, \ref{fig:FeCrStructAniz2}, and the corresponding transmissivities in Figs.~\ref{fig:FeCrStructAnizTransm}, \ref{fig:FeCrStructAnizTransm2}. We can see that for small repetition numbers, i.e., not too thick samples, we get the same results using both methods for each polarisation setting. For larger thicknesses, numerical instabilities arise with the transfer matrix method around the total reflection region; for higher angles, the results obtained by the two methods are in good agreement.

\begin{figure*}[htb]
\includegraphics[width=7.5cm, clip]{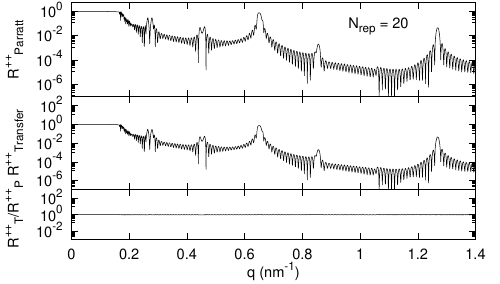}~
\includegraphics[width=7.5cm, clip]{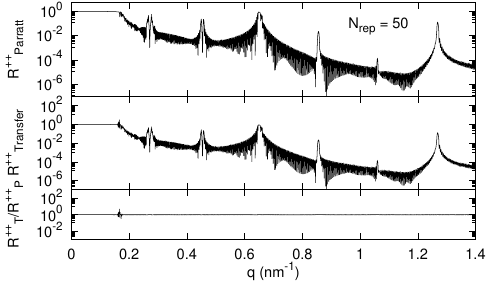}
\includegraphics[width=7.5cm, clip]{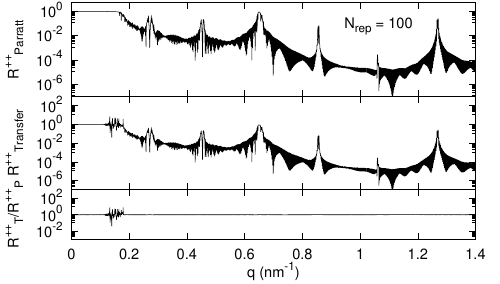}~
\includegraphics[width=7.5cm, clip]{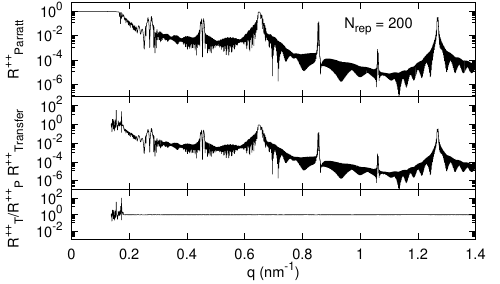}
\caption{Comparison of polarised neutron reflectometry results for [Cr(4 nm)/Fe(6 nm)/Cr(4 nm)/Fe(6 nm)/Cr(4 nm)/Fe(6 nm)]\(_{N_\text{rep}}\)/MgO multilayer systems for different hexalayer repetition numbers \(N_\text{rep}.\) In the hexalayers, the iron layers are magnetised in different directions (30°, 120°, 200° measured from the plane of incidence) in the sample plane. Here, the interfaces have a roughness of 0.5 nm and 0.4 nm on the top of Cr and Fe layers, respectively. The reflectivities for the ++ polarizer-analyzer state were computed using both the transfer matrix method \(\left(R_\text{Transfer}\right)\) and the generalized Parratt algorithm \(\left(R_\text{Parratt}\right)\) described here. The ratio of these results \(\left(R_\text{T}/R_\text{P} = R_\text{Transfer}/R_\text{Parratt}\right)\) is also plotted. The results of numerical instabilities are apparent in and near the total reflection region.}\label{fig:FeCrStructAnizRough}
\end{figure*}
One limitation of the method presented here is that, in the anisotropic case, no analytical approximation is available to account for interface (and/or magnetic) roughness. In such situations, the `brute force method' may be required, in which the rough interface is replaced by an equivalent system of thin layers whose scattering length density (or susceptibility in the case of X-rays) and effective magnetic induction vary according to the Gaussian profile commonly assumed for interfaces.
Thus, we approximate the depth profile of these quantities with a step function. This method is also known in the literature by the names of \textit{`graded-interface approach'} \cite{Esashi-2021} and \textit{multi-slicing} \cite{Macke-2014}. (This approach has one drawback: the calculation may get 1-2 orders of magnitude slower.) We used such an algorithm for the calculation of the [Cr(4 nm)/Fe(6 nm)/Cr(4 nm)/Fe(6 nm)/Cr(4 nm)/Fe(6 nm)]\(_{N_\text{rep}}\)/MgO profile, where we added roughness to the interfaces. 0.5 nm and 0.4 nm rms roughness on the top of Cr and Fe layers, respectively. The results are plotted in Fig.~\ref{fig:FeCrStructAnizRough}. The Parratt algorithm is obviously stable as well, and the methods give the same results in the `regions of stability'.

\begin{figure*}[htb]
\includegraphics[width=7.5cm, clip]{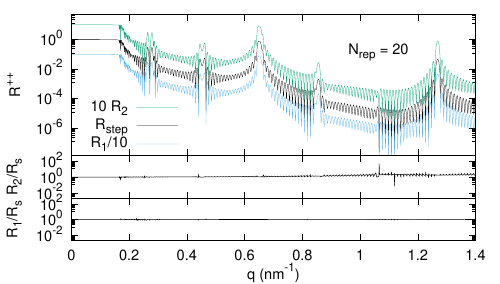}~
\includegraphics[width=7.5cm, clip]{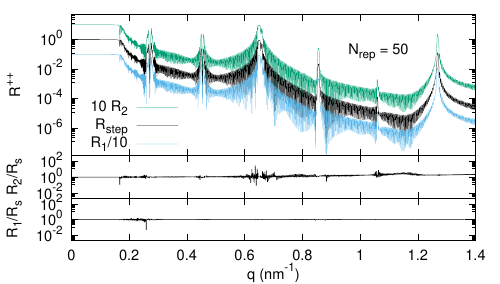}
\includegraphics[width=7.5cm, clip]{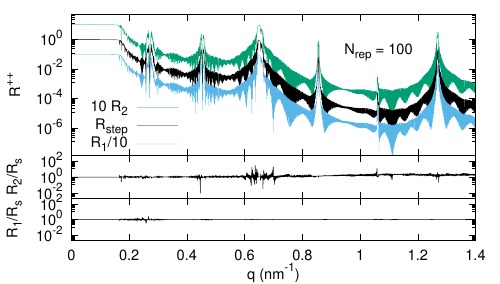}~
\includegraphics[width=7.5cm, clip]{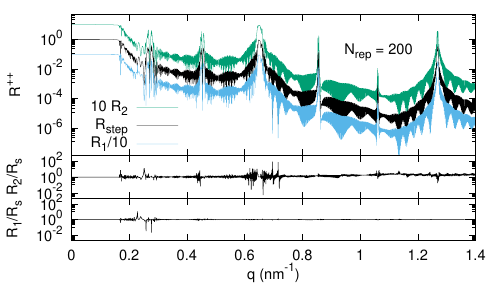}
\caption{Comparison of polarised neutron reflectometry results of multilayer systems with rough interfaces. It compares the results of the brute force method, which approximates the interface profile with a step function, with the two analytical approximations. The system is the same as in Fig.~\ref{fig:FeCrStructAnizRough}. The reflectivities for the ++ polarizer-analyzer state were computed using the generalized Parratt algorithm described here. The ratio of these results \(\left(R_\text{1,2}/R_\text{s} = R_\text{1,2 analytical approximation}/R_\text{step function}\right)\) is also plotted.}\label{fig:FeCrStructRoughnessStepFunctAnalyt}
\end{figure*}
We carried out calculations to compare the analytical approximations that account for interface roughness with results obtained using the `brute force method', in which the interface profile is modeled as a step function profile. The results are shown in Fig.~\ref{fig:FeCrStructRoughnessStepFunctAnalyt}. The second analytical approach exhibits significantly larger discrepancies, particularly at higher angles (larger \(q\)) and for larger repetition numbers. The first method shows difficulties near the total reflection edge and the first Bragg peak for high repetition numbers; therefore, it is advisable to validate its results by comparison with the `brute force method'.

The speed of the transfer matrix method and the generalized Parratt method is in the same order of magnitude in the general case. If we have repetition layer groups, the transfer matrix method may be faster, as we calculate the characteristic matrix of that group only once and use the \(N\)-th power of this matrix, where \(N\) is the number of repetitions, which speeds up the calculation. We do not have such a shortcut in the case of the Parratt methods, as (\ref{eq:ParrattGenForm}) does not allow such easy `parallelization'. If we have such repetitions, we can obtain some speed up by storing the Fresnel coefficient matrices and \(\tens{P}^+_l\) to avoid their recalculation.

The generalized Parratt algorithm, in its current formulation, is well-suited for computing the wave function as a function of depth within the sample. The use of the \(\tenscr{R}_{l}\) and \(\tenscr{T}_{l}\) matrices allows the calculation of all amplitudes \(\irottv{A}^{\pm}_{l},\) and the method can therefore be extended to off-specular \cite{Sinha-1988,DeakOffsp07} calculations, in which numerical instabilities may also pose challenges \cite{Hafner-2021,VeresOffspek2017}.

\section{Conclusion}\label{sec:Conclusion}

We have presented a numerically stable generalized Parratt method for anisotropic media, which can serve as an alternative to the characteristic matrix method, known to be prone to numerical instabilities.
In our opinion, this generalized method is useful for polarized neutron reflectometry, (nuclear) resonant x-ray (M\"{o}ssbauer) reflectometry.
We have shown that it can also be used in transmissivity calculations.
Future application to off-specular problems is possible, but that requires further investigation.

We demonstrated the capabilities of the new method by comparing the calculated results obtained using the method of characteristic matrices \cite{Deak96,Deak01},
implemented in the programs EFFI \cite{Spiering00} and FitSuite \cite{Sajti2009}. The new method has been implemented in the latest version of the latter, readily available for X-ray and neutron reflectometry of
layered systems of arbitrary complexity. The problem of interface roughness has also been examined; we showed different approximations, which we contrasted using test calculations.
\begin{dataavailability}
The files containing the calculated data used to create the figures are deposited in HUN-REN ARP data repository \cite{SajtiDeak-2026ParrattAdat}.
\end{dataavailability}
\bibliographystyle{apsrev}
\bibliography{
references}

\appendix
\section{Derivation of the generalized Parratt formulae: (\ref{eq:ParrattGenForm}) and (\ref{eq:TransmissIteration})}\label{app:DerivParratt}
It can be seen, that  according to (
\ref{eq:PplmiDef}, \ref{eq:ModusAmplTrMatr}, \ref{eq:Wmatr}, \ref{eq:WmatrInv0}),
\begin{equation}
\begin{split}
\tens{Z}_{l} &= \tens{W}_{l+1}^{-1}(0) \tens{W}_{l}(0) \tenscr{P}_{l}(d_l) =
\left(\begin{array}{cc}\tens{\Delta}^+_l & \tens{\Delta}^-_l\\ \tens{\Delta}^-_l & \tens{\Delta}^+_l
 \end{array}\right) \tenscr{P}_{l}(d_l)\\
 &=
\left(\begin{array}{cc}\tens{\Delta}^+_l \tenscr{P}_l^+(d_l) & \tens{\Delta}^-_l \tenscr{P}_l^-(d_l)\\ \tens{\Delta}^-_l \tenscr{P}_l^+(d_l) & \tens{\Delta}^+_l \tenscr{P}_l^-(d_l)
 \end{array}\right)
 \label{eq:ZPDelta2},
 \end{split}
\end{equation}
where
\begin{equation}
\tens{\Delta}^{\pm}_l
=\frac{1}{2}\tens{U}_{l+1}^{-1}\left[ \tens{I} \pm \tens{\kappa}_{l+1}^{-1}\tens{\kappa}_{l} \right] \tens{U}_{l} =
\frac{1}{2}\tens{U}_{l+1}^{-1} \tens{\kappa}_{l+1}^{-1} \left[ \tens{\kappa}_{l+1} \pm \tens{\kappa}_{l} \right] \tens{U}_{l}, \label{eq:DeltaDef}
\end{equation}
hence omitting the \(d_l\) arguments of \(\tenscr{P}_l,\) we have
\begin{eqnarray}
{^l\tens{Z}_{[11]}} &=&\tens{\Delta}^+_l \tenscr{P}_l^+, \quad
{^l\tens{Z}_{[12]}} = \tens{\Delta}^-_l {\tenscr{P}_l^+}^{-1}, \nonumber\\
{^l\tens{Z}_{[21]}} &=& \tens{\Delta}^-_l \tenscr{P}_l^+,\quad
{^l\tens{Z}_{[22]}}^{-1} = \tenscr{P}_l^+ {\tens{\Delta}^+_l}^{-1}. \label{eq:ZMatrBlockDef}
\end{eqnarray}
Substituting this in (\ref{eq:ParrattWithZ}), we get
\begin{equation}
\begin{split}
&\tenscr{R}_{l} =
\left[\tens{I} - \tenscr{P}_l^+ {\tens{\Delta}^+_l}^{-1} \tenscr{R}_{l+1} {\tens{\Delta}^-_l} {\tenscr{P}_l^+}^{-1}  \right]^{-1}
\left[\tenscr{P}_l^+ {\tens{\Delta}^+_l}^{-1}\tenscr{R}_{l+1} {\tens{\Delta}^+_l}
\right.\\ &\left. \tenscr{P}_l^+ - \tenscr{P}_l^+ {\tens{\Delta}^+_l}^{-1} {\tens{\Delta}^-_l}\tenscr{P}_l^+ \right] 
=
\tenscr{P}_l^+ \left[\tens{I} - {\tens{\Delta}^+_l}^{-1} \tenscr{R}_{l+1} {\tens{\Delta}^-_l}\right]^{-1} \\
&{\tenscr{P}_l^+}^{-1}
\tenscr{P}_l^+\left[{\tens{\Delta}^+_l}^{-1} \tenscr{R}_{l+1} {\tens{\Delta}^+_l} \tenscr{P}_l^+ -  {\tens{\Delta}^+_l}^{-1} {\tens{\Delta}^-_l}\tenscr{P}_l^+ \right] =
 \tenscr{P}_l^+\\
 &\left[\tens{I} - {\tens{\Delta}^+_l}^{-1} \tenscr{R}_{l+1} {\tens{\Delta}^-_l} \right]^{-1}  \left[{\tens{\Delta}^+_l}^{-1} \tenscr{R}_{l+1} {\tens{\Delta}^+_l}  -  {\tens{\Delta}^+_l}^{-1} {\tens{\Delta}^-_l} \right]
 \tenscr{P}_l^+. \label{eq:RDelta1}
\end{split}
\end{equation}

Using (\ref{eq:DeltaDef}) and (\ref{eq:FresnelRT}), we may write
\begin{equation}
\begin{split}
&\tens{\Delta}^+_l = \tens{U}_{l+1}^{-1} \tenscr{D}_{l+1,l}^{-1} \tens{U}_{l},\quad
\tens{\Delta}^-_l = -\tens{U}_{l+1}^{-1} \tenscr{D}_{l+1,l}^{-1}\tenscr{F}_{l,l+1} \tens{U}_{l},\\
&{\tens{\Delta}^+_l}^{-1} = \tens{U}_{l}^{-1} \tenscr{D}_{l+1,l} \tens{U}_{l+1},\quad
{\tens{\Delta}^+_l}^{-1} \tens{\Delta}^-_l = - \tens{U}_{l}^{-1} \tenscr{F}_{l,l+1} \tens{U}_{l}.\label{eq:DeltaDefApp}
\end{split}
\end{equation}
We can rewrite (\ref{eq:RDelta1}) using these expressions, like
\begin{equation}
\begin{split}
&\tenscr{R}_{l} =
\tenscr{P}_l^+
\left(\tens{I} +\tens{U}_{l}^{-1} \tenscr{D}_{l+1,l} \tens{U}_{l+1} \tenscr{R}_{l+1} \tens{U}_{l+1}^{-1} \tenscr{D}_{l+1,l}^{-1}\tenscr{F}_{l,l+1} \tens{U}_{l}   \right)^{-1}\\
&\left( \tens{U}_{l}^{-1} \tenscr{D}_{l+1,l} \tens{U}_{l+1} \tenscr{R}_{l+1} \tens{U}_{l+1}^{-1} \tenscr{D}_{l+1,l}^{-1} \tens{U}_{l}  +\tens{U}_{l}^{-1} \tenscr{F}_{l,l+1} \tens{U}_{l} \right)
\tenscr{P}_l^+.
\end{split}
\end{equation}

From this expression, we obtain the generalized Parratt formula using the definitions (\ref{eq:ReflHatAmplRefl}) and (\ref{eq:PeqUPU}),
\begin{eqnarray}
\tenshat{R}_{l} &= \tens{P}_l^+
\left(
\tens{I} +
\tenscr{D}_{l+1,l}\tenshat{R}_{l+1}\tenscr{D}_{l+1,l}^{-1} \tenscr{F}_{l,l+1}
\right)^{-1}\nonumber\\
&\left(
\tenscr{F}_{l,l+1} +
\tenscr{D}_{l+1,l} \tenshat{R}_{l+1}\tenscr{D}_{l+1,l}^{-1}
\right)
\tens{P}_l^+. \label{eq:RParrattAppendix}
\end{eqnarray}

The matrix characterizing the wave transmission through the bulk of \(l\)-th layer and through the \((l, l+1)\)-th interface according to (\ref{eq:TransmissionStart}) is
\begin{equation}
\begin{split}
\tens{T}_{l} &= \tens{U}_{l+1}{^l\tens{Z}_{[11]}} \tens{U}_{l}^{-1} + \tens{U}_{l+1} {^l\tens{Z}_{[12]}}  \tens{U}_{l}^{-1} \tenshat{R}_{l} = \\
&\tens{U}_{l+1}\tens{\Delta}^+_l \tenscr{P}_l^+ \tens{U}_{l}^{-1} +
\tens{U}_{l+1} \tens{\Delta}^-_l {\tenscr{P}_l^+}^{-1}  \tens{U}_{l}^{-1} \tenshat{R}_{l} =\\
&\tens{U}_{l+1}\tens{\Delta}^+_l \tens{U}_{l}^{-1} \tens{P}_l^+ + \tens{U}_{l+1} \tens{\Delta}^-_l \tens{U}_{l}^{-1} {\tens{P}_l^+}^{-1} \tenshat{R}_{l}. \label{eq:TransmissWidthDelta}
\end{split}
\end{equation}
This equation, after substitution of (\ref{eq:ZMatrBlockDef}), (\ref{eq:DeltaDefApp}), and using (\ref{eq:PeqUPU}) becomes
\begin{equation}
\tens{T}_{l} = \tenscr{D}_{l+1,l}^{-1} \tens{P}_l^+ - \tenscr{D}_{l+1,l}^{-1}\tenscr{F}_{l,l+1} {\tens{P}_l^+}^{-1} \tenshat{R}_{l}.
\end{equation}
Using (\ref{eq:RParrattAppendix}), we get the final formula for the transmission
\begin{eqnarray}
\tens{T}_{l} &= \tenscr{D}_{l+1,l}^{-1} \left[
\tens{I} - \tenscr{F}_{l,l+1}
\left(
\tens{I} +
\tenscr{D}_{l+1,l}\tenshat{R}_{l+1}\tenscr{D}_{l+1,l}^{-1} \tenscr{F}_{l,l+1}
\right)^{-1} \right. \nonumber\\
&\left.
\left(
\tenscr{F}_{l,l+1} +
\tenscr{D}_{l+1,l} \tenshat{R}_{l+1}\tenscr{D}_{l+1,l}^{-1}
\right)
\right] \tens{P}_l^+.
\end{eqnarray}

\section{Derivation of the generalized Parratt formula with roughness}\label{app:DerivRoughParratt}
Starting from (\ref{eq:ZRoughStartAver}) and (\ref{eq:ZPDelta2}),

\begin{widetext}
\begin{equation}
\tens{Z}^\text{rough}_{l} = \left(\begin{array}{cc}
\left<\tenscr{P}_{l+1}^-(\Delta z_{l+1}) \tens{\Delta}^+_l  \tenscr{P}_{l}^+(\Delta z_{l+1}) \right> &
\left<\tenscr{P}_{l+1}^-(\Delta z_{l+1}) \tens{\Delta}^-_l  \tenscr{P}_{l}^-(\Delta z_{l+1}) \right> \\
\left<\tenscr{P}_{l+1}^+(\Delta z_{l+1}) \tens{\Delta}^-_l  \tenscr{P}_{l}^+(\Delta z_{l+1}) \right> &
\left<\tenscr{P}_{l+1}^+(\Delta z_{l+1}) \tens{\Delta}^+_l  \tenscr{P}_{l}^-(\Delta z_{l+1}) \right>
\end{array}
\right) \tenscr{P}_l(\d_l).
\end{equation}
\begin{equation}
\left<\tenscr{P}_{l+1}^\mp(\Delta z_{l+1}) \tens{\Delta}^+_l  \tenscr{P}_{l}^\pm(\Delta z_{l+1}) \right>
= \left(\begin{array}{cc}
\left<e^{\mp\imunit(\kappa_{l+1;1} - \kappa_{l;1}) \Delta z_{l+1}} \right> &
\left<e^{\mp\imunit(\kappa_{l+1;1} - \kappa_{l;2}) \Delta z_{l+1}} \right>\\
\left<e^{\mp\imunit(\kappa_{l+1;2} - \kappa_{l;1}) \Delta z_{l+1}} \right> &
\left<e^{\mp\imunit(\kappa_{l+1;2} - \kappa_{l;2}) \Delta z_{l+1}} \right>
\end{array}\right)
\odot\tens{\Delta}^+_l,
\end{equation}

\begin{equation}
\left<\tenscr{P}_{l+1}^\pm(\Delta z_{l+1}) \tens{\Delta}^-_l  \tenscr{P}_{l}^\pm(\Delta z_{l+1}) \right>
= \left(\begin{array}{cc}
\left<e^{\pm\imunit(\kappa_{l+1;1} + \kappa_{l;1}) \Delta z_{l+1}} \right> &
\left<e^{\pm\imunit(\kappa_{l+1;1} + \kappa_{l;2}) \Delta z_{l+1}} \right>\\
\left<e^{\pm\imunit(\kappa_{l+1;2} + \kappa_{l;1}) \Delta z_{l+1}} \right> &
\left<e^{\pm\imunit(\kappa_{l+1;2} + \kappa_{l;2}) \Delta z_{l+1}} \right>
\end{array}\right)
\odot\tens{\Delta}^-_l.
\end{equation}
\end{widetext}
From this, it is apparent that we have to calculate the averages of the form \(\left<e^{\pm\imunit \kappa z} \right>.\)
Assuming that the interface is the result of an ergodic random process and the height probability distribution is Gaussian with standard deviation \(\sigma,\) which denotes the rms roughness of the interface in depth \(z,\) we have
\begin{equation}
\left<e^{\pm\imunit \kappa z} \right> = \int e^{\pm\imunit \kappa z} \frac{1}{\sqrt{2\pi} \sigma} \e^{-\frac{z^2}{2\sigma^2}} \d z = e^{-\frac{1}{2}\sigma^2\kappa^2}.
\end{equation}
Substituting this in the matrices above replacing \(\sigma\) with \(\sigma_l\),  we will get the matrices \(\tens{G}^\pm\) defined in   (\ref{eq:GDef}).

We should mention an important theorem of the Hadamard product
\begin{equation}
\tens{D} \left(\tens{A} \odot \tens{B}\right) \tens{E} = \left(\tens{D} \tens{A} \tens{E}\right) \odot \tens{B}
= \left(\tens{A} \tens{E}\right) \odot \left(\tens{D} \tens{B}\right) = \tens{A} \odot \left(\tens{D} \tens{B} \tens{E}\right), \label{eq:HadamardProdDiagProp}
\end{equation}
if \(\tens{D}\) and \(\tens{E}\) are diagonal matrices \cite{Horn-b2008}. 
This is relevant in (\ref{eq:ZRoughCorr}), as \(\tenscr{P}\) is diagonal, therefore \(\left(\tens{G}\odot\tens{\Delta}\right) \tenscr{P} = \left(\tens{G}\tenscr{P}\right)\odot\tens{\Delta}  = \tens{G}\odot\left(\tens{\Delta} \tenscr{P}\right).\)

Taking into account the interface roughness from (\ref{eq:ZRoughCorr}), (\ref{eq:GDef}), (\ref{eq:HadamardProdDiagProp}), and (\ref{eq:ZPDelta2}), it is apparent that
\begin{equation}
\tens{Z}^\text{rough}_{l} = \left(\begin{array}{cc}\tens{G}^-_l\odot\tens{\Delta}^+_l \tenscr{P}_l^+(d_l) & \tens{G}^+_l\odot\tens{\Delta}^-_l \tenscr{P}_l^-(d_l)\\ \tens{G}^+_l\odot\tens{\Delta}^-_l \tenscr{P}_l^+(d_l) & \tens{G}^-_l\odot\tens{\Delta}^+_l \tenscr{P}_l^-(d_l)
 \end{array}\right)
\end{equation}
As this has the same form as (\ref{eq:ZPDelta2}) if we replace \(\tens{\Delta}^\pm_l\) with
\begin{equation}
\tens{\epsilon}^{\pm}_l =  \tens{G}^\mp_l \odot \tens{\Delta}^{\pm}_l = \tens{\Delta}^{\pm}_l \odot \tens{G}^\mp_l, \label{eq:epsilonDefApp}
\end{equation}
we may get the same way the reflectivity using the same substitution in (\ref{eq:RDelta1})

\begin{eqnarray}
\tenscr{R}_{l} &=&
\tenscr{P}_l^+
\left(\tens{I} - {\tens{\epsilon}^+_l}^{-1} \tenscr{R}_{l+1} {\tens{\epsilon}^-_l}   \right)^{-1} \nonumber\\
&&\left( {\tens{\epsilon}^+_l}^{-1} \tenscr{R}_{l+1} {\tens{\epsilon}^+_l}  -  {\tens{\epsilon}^+_l}^{-1} {\tens{\epsilon}^-_l} \right)
 \tenscr{P}_l^+.
\end{eqnarray}
 Similarly, introducing
\begin{equation}
\tens{\delta}^{\pm}_l = \tens{U}_{l+1} \tens{\epsilon}^{\pm}_l \tens{U}_{l}^{-1} = \tens{U}_{l+1} \left(\tens{G}^\mp_l \odot \tens{\Delta}^{\pm}_l\right) \tens{U}_{l}^{-1}, \label{eq:deltaDefApp}
\end{equation}
we may obtain the expression
\begin{eqnarray*}
\tenshat{R}_{l}&=&
\tens{P}_l^+ \left(\tens{I} - {\tens{\delta}^+_l}^{-1} \tenshat{R}_{l+1} {\tens{\delta}^-_l} \right)^{-1}\\
 &&\left( {\tens{\delta}^+_l}^{-1} \tenshat{R}_{l+1} {\tens{\delta}^+_l}  -  {\tens{\delta}^+_l}^{-1} {\tens{\delta}^-_l} \right)
 \tens{P}_l^+.
\end{eqnarray*}
From this, using (\ref{eq:deltaDefApp}), (\ref{eq:epsilonDefApp}), and (\ref{eq:DeltaDefApp}), we get (\ref{eq:RoughParrattGenForm}).

For the transmission substituting in (\ref{eq:TransmissWidthDelta}), we get
\begin{eqnarray*}
&\tens{T}_{l} =\tens{\delta}^+_l \tens{P}_l^+ +  \tens{\delta}^-_l {\tens{P}_l^+}^{-1} \tenshat{R}_{l} =
\left[ \tens{\delta}^+_l + \tens{\delta}^-_l
\left(\tens{I} - {\tens{\delta}^+_l}^{-1} \right. \right.\nonumber\\
&\left. \left.
\tenshat{R}_{l+1} {\tens{\delta}^-_l} \right)^{-1}
 \left( {\tens{\delta}^+_l}^{-1} \tenshat{R}_{l+1} {\tens{\delta}^+_l}  -  {\tens{\delta}^+_l}^{-1} {\tens{\delta}^-_l} \right)
\right] \tens{P}_l^+.
\end{eqnarray*}

\section{Alternative derivation of the generalized Parratt formula with roughness, following Röhlsberger's approach}\label{app:Rohlsberger}
According to Röhlsberger \cite{RohlsbergerHypInt99,Rohlsberger-b2009-chp4}, a rough interface can be taken into account by replacing the product of the transfer matrices of the form
\begin{eqnarray}
&\e^{\imunit \tens{H}_{l+1} d_{l+1}} \e^{\imunit \tens{H}_l d_l}
\quad \text{with}\quad
\e^{\imunit \tens{H}_{l+1} d_{l+1}} 
\nonumber\\
&\e^{\tens{Q}_{l,l+1}}
\e^{\imunit \tens{H}_l d_l},\quad \tens{Q}_{l,l+1} = \sum\limits_{n=1}^\infty \frac{1}{n!} \left(\frac{\sigma_{l,l+1}^2}{2}\right)^n \left[\tens{H}_l,\tens{H}_{l+1}\right]_{2n},\nonumber
\end{eqnarray}
where \(d_l\) denotes the layer thicknesses, \(\sigma_{l,l+1}\) is the interface roughness, the \(4\times 4\) matrix \(\tens{H_l}\) depends on the material parameters of the layer, and the wavenumber of the incident beam, \([A,B]_2 = [A,B]\) denotes the commutator, and \([A,B]_{2n+2} = [[[A,B]_{2n},A],B]\) denotes higher order commutators.
In our case, we can see that this transfer matrix is of the form \(\tens{W}_l(0)\tenscr{P}_l(d_l)\tens{W}_l^{-1}(0)\) according to (\ref{eq:WaveFuncTransfMatr}), from which, using 
(\ref{eq:PplmiDef}), (\ref{eq:kappaMatrDef}), (\ref{eq:Wmatr}) and (\ref{eq:WmatrInv0}) we obtain
\begin{equation*}
\tens{H}_l=\tens{W}_l(0)\!\left[\begin{array}{cc}
\tens{U}_l^{-1}\tens{\kappa}_l\tens{U}_l &\\
&-\tens{U}_l^{-1}\tens{\kappa}_l \tens{U}_l
\end{array} \right]\!\tens{W}_l^{-1}\!(0)
=  \imunit\!\left[
\begin{array}{cc}
&-\tens{I}\\
\tens{\kappa}_l^2
\end{array}
\right],
\end{equation*}
and it can be shown, that we can express the commutators in the following form, where the \(\tens{c}_n\) matrices are defined by a recurrence relation
\begin{eqnarray}
\left[\tens{H}_l, \tens{H}_{l+1} \right]_{2n} &=
\left(
\begin{array}{cc}
\tens{c}_n&\\ %
&-\tens{c}_n
\end{array}
\right), \quad \tens{c}_1 = \tens{\kappa}_{l+1}^2 - \tens{\kappa}_l^2,
\\
\tens{c}_{n+1} &= - \left(2\tens{c}_n \tens{\kappa}_{l+1}^2 + \tens{c}_n \tens{\kappa}_l^2 + \tens{\kappa}_l^2 \tens{c}_n\right).
\end{eqnarray}
Thus, \(\tens{Q}_{[22]} = - \tens{Q}_{[11]},\) and therefore \(\tens{E}_{[22]} = {\tens{E}_{[11]}}^{-1},\) where \(\tens{E} = \exp \tens{Q}.\)
The corrected amplitude transfer matrix will be
\begin{eqnarray}
^{\text{corr}}\tens{Z}_{l} &= \tens{W}_{l+1}^{-1}(0)  \e^{\tens{Q}_{l,l+1}} \tens{W}_{l}(0) \tenscr{P}_l(d_l) = \nonumber\\
&\left(\begin{array}{cc}\tens{\Delta^\prime}^+_l \tenscr{P}_l^+(d_l) & \tens{\Delta^\prime}^-_l \tenscr{P}_l^-(d_l)\\ \tens{\Delta^\prime}^-_l \tenscr{P}_l^+(d_l) & \tens{\Delta^\prime}^+_l \tenscr{P}_l^-(d_l)
 \end{array}\right), \label{eq:ZPDeltaRohlsberger}
\end{eqnarray}
where
\begin{eqnarray*}
&\tens{\Delta^\prime}^{\pm}_l
=\frac{1}{2}\tens{U}_{l+1}^{-1}\left[ \tens{\eta}_{l,l+1} \pm \tens{\kappa}_{l+1}^{-1}\tens{\eta}_{l,l+1}^{-1}\tens{\kappa}_{l} \right] \tens{U}_{l} =
\tens{U}_{l+1}^{-1} \frac{1}{2}\tens{\kappa}_{l+1}^{-1} \nonumber\\
&\left[ \tens{\kappa}_{l+1}\tens{\eta}_{l,l+1} \pm \tens{\eta}_{l,l+1}^{-1}\tens{\kappa}_{l} \right] \tens{U}_{l},
\quad\text{and}\quad \tens{\eta}_{l,l+1} = \e^{{\tens{Q}_{l,l+1}}_{[11]}}.
\label{eq:DeltaPrDef}
\end{eqnarray*}
The matrices in (\ref{eq:ZPDelta2}) and (\ref{eq:ZPDeltaRohlsberger}) have obviously similar structures. Let us introduce the matrices
\begin{eqnarray*}
\!&\!^{R}\tenscr{F}_{l,l+1}\!=\!\left[\tens{\eta}_{l,l+1}^{-1} \tens{\kappa}_{l} + \tens{\kappa}_{l+1} \tens{\eta}_{l,l+1}\right]^{-1}\!
 \left[\tens{\eta}_{l,l+1}^{-1} \tens{\kappa}_{l} - \tens{\kappa}_{l+1}\tens{\eta}_{l,l+1} \right], \nonumber\\
\!&\!^{R}\tenscr{D}_{l,l+1}\!=\!2 \left[\tens{\eta}_{l,l+1}^{-1}\tens{\kappa}_{l} + \tens{\kappa}_{l+1} \tens{\eta}_{l,l+1} \right]^{-1} \tens{\kappa}_{l},
\end{eqnarray*}
which in the case of no roughness \((\sigma^2_{l,l+1} = 0, \tens{\eta}_{l,l+1} =\tens{I})\) will give back the Fresnel-coefficients defined in (\ref{eq:FresnelRT}), thus they can be interpreted as the corrected coefficient matrices for rough interfaces. As the expressions
\begin{eqnarray*}
&\tens{\Delta^\prime}^+_l = \tens{U}_{l+1}^{-1} {^{R}}\tenscr{D}_{l+1,l}^{-1} \tens{U}_{l},\Nquad
\tens{\Delta^\prime}^-_l = -\tens{U}_{l+1}^{-1} {^{R}}\tenscr{D}_{l+1,l}^{-1} {^{R}}\tenscr{F}_{l,l+1} \tens{U}_{l},\\
&{\tens{\Delta^\prime}^+_l}^{-1} = \tens{U}_{l}^{-1} {^{R}}\tenscr{D}_{l+1,l} \tens{U}_{l+1},\Nquad
{\tens{\Delta^\prime}^+_l}^{-1} \tens{\Delta^\prime}^-_l = - \tens{U}_{l}^{-1} {^{R}}\tenscr{F}_{l,l+1} \tens{U}_{l}.
\end{eqnarray*}
and (\ref{eq:DeltaDefApp}) are of the same form, we may use the iteration formulae (\ref{eq:ParrattGenForm}) and (\ref{eq:TransmissIteration}), just replacing the Fresnel matrices with their corrected form.

\section{The derivation of generalized Parratt formula (\ref{deakParratt})} \label{app:DeakParratt}
Starting from equations (\ref{eq:LDeak}) and (\ref{parratt_base}), and using the notations \(\tens{\delta}_l = \imunit k d_l \tens{\gamma}_l \sin \theta,\) \(\tens{c}_l = \cosh \tens{\delta}_l,\) \(\tens{s}_l = \sinh \tens{\delta}_l,\) and \(\tens{R}_{l}^\pm = \left(\tens{I}\pm\tens{R}_{l}\right)\)
\begin{eqnarray*}
\tens{R}_{l}^-   \Psi_{l}^{0} &=
\tens{\gamma}_{l}^{-1}
\left[
\tens{c}_l \tens{\gamma}_{l-1}\tens{R}_{l-1}^-
+ \tens{\gamma_l} \tens{s}_l \tens{R}_{l-1}^+
\right] \Psi_{l-1}^{0} \\
\tens{R}_{l}^+    \Psi_{l}^{0} &=
\left[
\tens{\gamma_l}^{-1} \tens{s}_l \tens{\gamma}_{l-1}\tens{R}_{l-1}^-
+ \tens{c}_l \tens{R}_{l-1}^+
\right] \Psi_{l-1}^{0}
 \label{parratt_base2}
\end{eqnarray*}
As \(\tens{\gamma}_l\) and its inverse commute, furthermore, as it commutes with its functions like \(\tens{c}_l\) and \(\tens{s}_l,\) we may write
\begin{eqnarray*}
\tens{R}_{l}^-   \Psi_{l}^{0} &=
\left[
\tens{c}_l \tens{\gamma}_{l}^{-1}\tens{\gamma}_{l-1}\tens{R}_{l-1}^-
+ \tens{s}_l \tens{R}_{l-1}^+
\right] \Psi_{l-1}^{0} \nonumber\\
\tens{R}_{l}^+    \Psi_{l}^{0} &=
\left[
\tens{s}_l \tens{\gamma_l}^{-1}\tens{\gamma}_{l-1}\tens{R}_{l-1}^-
+ \tens{c}_l \tens{R}_{l-1}^+
\right] \Psi_{l-1}^{0}
\label{parratt_base3}
\end{eqnarray*}

Summing and subtracting the equations above from each other, we get
\begin{eqnarray}
&2 \Psi_{l}^{0} =
\left(\tens{c}_l + \tens{s}_l \right)\left[
 \tens{\gamma}_{l}^{-1}\tens{\gamma}_{l-1}\tens{R}_{l-1}^-
+ \tens{R}_{l-1}^+
\right] \Psi_{l-1}^{0} \label{parratt_base4a}\\
&2 \tens{R}_{l}  \Psi_{l}^{0} =
\left(\tens{c}_l - \tens{s}_l \right) \left[
- \tens{\gamma_l}^{-1}\tens{\gamma}_{l-1}\tens{R}_{l-1}^-
+  \tens{R}_{l-1}^+
\right] \Psi_{l-1}^{0}
\label{parratt_base4b}
\end{eqnarray}
Substituting (\ref{parratt_base4a}) into (\ref{parratt_base4b}), and applying the identities \(\e^{\pm x} = \cosh x \pm \sinh x,\) and reordering the expressions, we get
\begin{eqnarray*}
& \tens{R}_{l} \e^{\tens{\delta}_l}\left[
 \tens{\gamma}_{l}^{-1}\tens{\gamma}_{l-1}\tens{R}_{l-1}^-
+ \tens{R}_{l-1}^+
\right] =
 \tens{R}_{l} \e^{\tens{\delta}_l}\left[
 \left(\tens{\gamma}_{l}^{-1}\tens{\gamma}_{l-1} + \tens{I}\right) + \right.\\
 &\left.
 \left(\tens{I} - \tens{\gamma}_{l}^{-1}\tens{\gamma}_{l-1} \right) \tens{R}_{l-1}
\right]
= \e^{-\tens{\delta}_l} \left[
- \tens{\gamma_l}^{-1}\tens{\gamma}_{l-1}\tens{R}_{l-1}^-
+ \tens{R}_{l-1}^+
\right] = \\
&\e^{-\tens{\delta}_l} \left[
\left(  \tens{I} - \tens{\gamma_l}^{-1}\tens{\gamma}_{l-1}\right)
+ \left(\tens{\gamma}_{l}^{-1}\tens{\gamma}_{l-1} + \tens{I}\right)\tens{R}_{l-1}
\right].
\end{eqnarray*}
Introducing the notation
\(\widehat{R}_{l} = \e^{\tens{\delta}_l} \tens{R}_{l} \e^{\tens{\delta}_l}\), and applying (\ref{aniFresnel_l}),
we have
\begin{eqnarray*}
&\tens{R}_{l-1} = \left[\left(\tens{\gamma}_{l}^{-1}\tens{\gamma}_{l-1} + \tens{I}\right) + \widehat{R}_l \left(\tens{\gamma}_{l}^{-1}\tens{\gamma}_{l-1} - \tens{I}\right)\right]^{-1}\\
&\left[\widehat{R}_l \left(\tens{\gamma}_{l}^{-1}\tens{\gamma}_{l-1} + \tens{I}\right)  + \left(\tens{\gamma}_{l}^{-1}\tens{\gamma}_{l-1} - \tens{I}\right) \right] = \\
& \left(\tens{\gamma}_{l}^{-1}\tens{\gamma}_{l-1} + \tens{I}\right)^{-1}
\left[\widehat{R}_l \tens{r}_{l} + \tens{I}\right]^{-1}
\left[\widehat{R}_l  + \tens{r}_{l}\right]
\left(\tens{\gamma}_{l}^{-1}\tens{\gamma}_{l-1} + \tens{I}\right).
\end{eqnarray*}
Using the expression
\begin{eqnarray*}
\tens{I} - \tens{r}_l &=& \tens{I} -\left[  \tens{\gamma}_{l-1}+\tens{\gamma}_{l}\right]^{-1}
\left(\tens{\gamma}_{l-1} -\tens{\gamma}_{l}\right) = \nonumber\\
&2& \left(\tens{\gamma}_{l-1}+\tens{\gamma}_{l}\right)^{-1} \tens{\gamma}_l = 2 \left(\tens{\gamma}_{l}^{-1}\tens{\gamma}_{l-1} + \tens{I}\right)^{-1},
\end{eqnarray*}
we get
\begin{equation*}
\tens{R}_{l-1} = \left(\tens{I} - \tens{r}_l\right)
\left[\widehat{R}_l \tens{r}_{l} + \tens{I}\right]^{-1}
\left[\widehat{R}_l  + \tens{r}_{l}\right]
\left(\tens{I} - \tens{r}_l\right)^{-1},
\end{equation*}
which is (\ref{deakParratt}).

\end{document}